\documentclass[11pt,reqno]{article}
\usepackage{amssymb, amsmath, amsthm, amsfonts, amscd, epsfig}
 \usepackage{color}
\usepackage[english]{babel}
\usepackage{bm}
\usepackage{graphicx, epsfig}
\usepackage{kotex,setspace}

\usepackage{subcaption}

\newcommand{\Be}{\mathbf{e}}
\newcommand{\Bj}{\mathbf{j}}
\newcommand{\Bt}{\mathbf{t}}
\newcommand{\Om}{\Omega}
\newcommand{\Bx}{\mathbf{x}}
\newcommand{\bU}{\mathbf{U}}
\newcommand{\bD}{\mathbf{D}}
\newcommand{\bF}{\mathbf{F}}
\newcommand{\bI}{\mathbf{I}}
\newcommand{\By}{\mathbf{y}}

\newcommand{\Bn}{\mathbf{n}}

\newcommand{\RR}{\mathbb{R}}
\newcommand{\CC}{\mathbb{C}}

\newcommand{\ZZ}{\mathbb{Z}}

\newcommand{\iu}{\mathrm{i}\mkern1mu}
\newcommand{\p}{\partial}

\newcommand{\pd}[2]{\frac {\p #1}{\p #2}}
\newcommand{\ds}{\displaystyle}
\newcommand{\eqnref}[1]{(\ref {#1})}

\newcommand{\beq}{\begin{equation}}
\newcommand{\eeq}{\end{equation}}
\newcommand{\RN}[1]{%
  \textup{\uppercase\expandafter{\romannumeral#1}}%
}

\newcommand{\tphi}{\widetilde{\varphi}}
\newcommand{\tn}{\widetilde{n}}

\numberwithin{equation}{section}
\numberwithin{figure}{section}

\begin{document}

\date{}

\title{Inclusions of general shapes having constant field inside the core and non-elliptical
neutral coated inclusions with anisotropic conductivity
}

\author{Mikyoung Lim\thanks{Department of Mathematical Sciences,
Korea Advanced Institute of Science and Technology, Daejeon 305-701, Korea (mklim@kaist.ac.kr).} 
\and Graeme W. Milton \thanks{Department of Mathematics, University of Utah, Salt Lake City, UT 84112, USA and Korea Advanced Institute of Science and Technology, Daejeon 34141, Korea (milton@math.utah.edu)}}

\maketitle

\begin{abstract}
For certain shapes of inclusions embedded in a body, the field inside the inclusion is uniform for some boundary condition. 
We provide a construction scheme for inclusions of general shapes having such a uniformity property in two dimensions based on the conformal mapping technique for the potential problem. 
Using this complex analysis method, we also design non-elliptical neutral coated inclusions with anisotropic conductivity. 
Neutral coated inclusions do not perturb a background uniform field when they are inserted into a homogeneous matrix. Although coated inclusions of various shapes are neutral to a single field, only concentric ellipses or confocal ellipsoids can be neutral to all uniform fields. 
This paper presents our work relating to the construction of non-elliptical coated inclusions with anisotropic conductivity in two dimensions that are neutral to all uniform fields, where the assignment of the flux condition on the boundary of the core depends on the applied background field. 
Using these neutral inclusions, we obtain cylindrical neutral inclusions in three dimensions, with no flux applied to the boundary of the core and with the anisotropic conductivity function of the shell given in accordance with the background uniform field.
\end{abstract}



\section{Introduction}

Most conducting (or dielectric or magnetic) objects inserted in a medium of constant conductivity (or permittivity or permeability)  in which there are uniform electric (or magnetic) fields have resulting fields that are generally
neither uniform inside nor outside the object. 
However, certain shapes of inclusions exist inside which the resulting field is uniform for an applied uniform loading. Poisson \cite{Poisson:1826:SMS} realized that the field inside an ellipsoid must be
uniform and explicit expressions for this field were obtained by Maxwell \cite{Maxwell:1954:TEMb}. Eshelby showed that an ellipse or an ellipsoid satisfies this uniformity property and conjectured the following: if an inclusion satisfies the uniformity property, then it should be an ellipse or an ellipsoid \cite{Eshelby:1957:DEF, Eshelby:1961:EII}. This conjecture was proved to be true within the class of simply connected domains \cite{Kang:2008:SPS, Liu:2008:SEC, Ru:1996:EIA, Sendeckyj:1970:EIP}. Following Liu et al. \cite{Liu:2008:preprint} and Liu \cite{Liu:2009:ECT, Liu:2010:SPE} (periodic structure), we denote an {\it $E$-inclusion} for an inclusion embedded in an infinite medium of constant conductivity (or embedded in a unit cell with periodic boundary conditions) 
that satisfies the Eshelby's uniformity property for at least one applied field. We also denote, following Kang et al. \cite{Kang:2011:SBV} and Bardsley et al. \cite{Bardsley:2017:CGB}, an {\it $E_\Omega$-inclusion} for an inclusion embedded in a body $\Omega$ of constant conductivity that satisfies the uniformity property for some appropriate boundary 
conditions on $\partial\Omega$. $E$-inclusions were investigated by Kang \cite{Kang:2009:CPS} and Liu \cite{Liu:2008:SEC} in relation to the classical Newtonian potential problem.
Finding $E$-inclusions in a unit cell with periodic boundary conditions is important for finding microstructures with extreme effective conductivity or
extreme effective bulk modulus that attain the Hashin-Shtrikman bounds or their anisotropic generalizations. Finding $E$- or $E_\Omega$-inclusions is also an important problem with practical applications for designing materials which for conductivity (or anti-plane elasticity) induce electric fields (or stresses) with small variances in the inclusion phase. These inclusions, which are tailored
to the applied field, are generally less likely to breakdown (or break) than inclusions with large variances of the electric fields (or stresses). 

A powerful technique for generating non-elliptical $E$- or $E_\Omega$-inclusions in two-dimensions
has been to use  hodographic transformations to solve the free boundary problem. Then the problem is reduced to a potential problem on a set of slits. This approach has been successfully used by  
Vigdergauz \cite{Vigdergauz:1976:IEI} to obtain periodic microstructures, known as Vigdergauz microstructures, which are two-dimensional $E$-inclusions with periodic boundary conditions (see also the work by Grabovsky and Kohn \cite{Grabovsky:1995:MMEb}). It has been extended to obtain two-dimensional periodic structures with multiple inclusions in the unit cell; see section 23.9 of \cite{Milton:2002:TC:book} and \cite{Antipov:2018:SMS}, and also for pairs of $E$-inclusions \cite{Cherepanov:1974:IPP, Kang:2008:IPS}. Additionally, it has been used to construct $E_\Omega$-inclusions
\cite{Kang:2011:SBV,Bardsley:2017:CGB}. The question arises as to whether this technique misses some inclusion shapes? In the context of the $E_\Omega$-inclusion problem we will
see that it does. Contrary to the analysis in \cite{Kang:2011:SBV,Bardsley:2017:CGB}, which suggested that only a limited family of simply connected shapes can be $E_\Omega$-inclusions, we will
see that any simply connected shape with an analytic boundary can be an $E_\Omega$-inclusion, for an appropriate choice of $\Omega$. Rather than using hodographic transformations, we will simply
use a conformal mapping that maps the region outside the inclusion to a region outside a circular disk and then solve the problem in the disk geometry using Laurent series.
The result shows that the hodographic approach has limitations.

We remark that an alternative  variational approach for obtaining $E$-inclusions was developed by Liu, James and Leo \cite{Liu:2007:PIM, Liu:2008:SEC}. Their approach is not limited to two-dimensions
and consequently they discovered three-dimensional periodic arrays of $E$-inclusions that saturate the Hashin-Shtrikman bounds \cite{Hashin:1962:VAT, Liu:2010:HSB} and they obtained $E$-inclusions
having disconnected components.

Our approach is quite similar to the conformal mapping method used in \cite{Milton:2001:NCI} to obtain neutral inclusions
which is the second subject of the paper. Some coated inclusions, when placed in a medium, do not disturb the exterior uniform field, and these are denoted as {\it neutral inclusions}.
They are in a sense invisible to the applied field \cite{Kerker:1975:IB}. 
Once a neutral inclusion has been found, similar inclusions, possibly of different sizes, can be added to the background matrix without altering the exterior uniform field \cite{Hashin:1962:EMH}. In this way it becomes possible to construct a composite, consisting of multiple inclusions and a background matrix, of which the effective property exactly coincides with that of the matrix. 
Two-dimensional conductivity problems can be equivalently considered as anti-plane elasticity problems. Well-known examples of neutral inclusions are assemblages of coated disks and spheres \cite{Hashin:1962:EMH, Hashin:1962:VAT}. As the field inside the core is uniform, these inclusions retain 
their neutrality even if the core material is made non-linear (see, for example, \cite{Hashin:1985:LIE,Jimenez:2013:NNI}).
Appropriately coated ellipses and ellipsoids, with the possibly anisotropic material parameters of the core, shell, and matrix, are neutral 
to all uniform fields \cite{Grabovsky:1995:MMEa,Kerker:1975:IB,Milton:2002:TC:book,Sihvola:1999:EMF,Sihvola:1997:DPI}, and these are the only shapes for which coated inclusions admit such a uniformity property \cite{Kang:2014:CIF,Kang:2016:OBV,Milton:2001:NCI}. The concept of neutral inclusion has been extensively studied, especially for designing an invisibility cloaking structure with metamaterials. For example, Zhou et al. constructed coated spheres or multi-layer spheres that are transparent to acoustic waves, elastic waves, or electromagnetic waves \cite{Zhou:2006:DEW,Zhou:2007:AWT,Zhou:2008:EWT}. Luo et al. \cite{Luo:2009:RAP} and Xi et al. \cite{Xi:2009:ODP} found neutral inclusions for the Helmholtz equation that were based on carpet cloaks. Later, Landy and Smith \cite{Landy:2013:FPU} physically realized these neutral inclusions with microwaves. Al\`{u} and Engheta \cite{Alu:2005:ATP} and Ammari et al. \cite{Ammari:2013:ENC} discovered multi-coated neutral inclusions for Maxwell's equations.

 Coated inclusions of non-elliptical shapes can be neutral to a single uniform field.
Milton and Serkov constructed various shapes of neutral inclusions in two dimensions with cores of perfectly conducting or insulating material by using the conformal mapping technique \cite{Milton:2001:NCI}, and Jarczyk and Mityushev extended this work to cores of finite conductivities \cite{Jarczyk:2012:NCI}. 
We refer readers to \cite{Milton:2002:TC:book} for more results and references. 
Recently, Kang and Li constructed weakly neutral inclusions of general shapes with imperfect interfaces \cite{Kang:2018:CWN:preprint}, and Choi et al. provided a numerical method to construct multi-coated neutral inclusions of general shapes \cite{Choi:2018:GME:preprint}. It is also worth mentioning that Kim and Lim discovered non-elliptical inclusion shapes such that with a suitable polynomial field at infinity, the field in the inclusion is uniform \cite{Kim:2018:EEC:preprint}.

In the present paper, we describe the construction of non-elliptical coated inclusions in two dimensions that are neutral to the uniform background fields of all directions, where the assignment of the flux condition on the boundary of the core depends on the applied background field. 
As the resulting {\it active neutral coated inclusions} are not detectable by outside measurements (with the given uniform applied field), one can view  this neutral inclusion problem with the flux condition as {\it active cloaking}
(see \cite{Vasquez:2009:AEC,Miller:2006:PC, Norris:2014:AEC, ONeill:2015:ACI,  Slvanayagam:2012:AEC} for other examples of active cloaking).
In addition, we design non-elliptical cylindrical neutral inclusions in three dimensions without imposing a flux on the boundary of the core, using the constructed two-dimensional neutral inclusions.
Our result for the three-dimensional neutral inclusion can be reinterpreted as a neutral inclusion in two dimensions in which currents are applied to the boundary of the core. 
In the special three-dimensional case where the shell has constant anisotropic conductivity, the condition for
neutrality forces the conductivity tensor of the shell to have an eigenvector aligned with the axis of the cylinder, and then the neutral inclusion shapes are exactly those found in a previous study \cite{Milton:2001:NCI}.

The remainder of this paper is organized as follows. 
In section \ref{sec:E_Omega} we describe the construction of $E_\Omega$-inclusions in two dimensions. Section \ref{sec:Neutral2D} is devoted to neutral inclusions with the active flux condition in two dimensions.
In section \ref{sec:3D} we consider the cylindrical neutral inclusion in three dimensions and reformulate the problem as a two-dimensional problem.
The paper ends with the conclusion.


\section{$E_\Omega$-inclusions of general shapes}\label{sec:E_Omega}
In this section, we present a new construction method for $E_\Om$-inclusions in two dimensions 
based on complex analysis.
Let $\Om$ and $D$ be simply connected bounded planar domains such that $\overline{D}\subset\Om$. The core $D$ has a constant, possibly anisotropic, conductivity $\bm{\sigma}_0$, and {it is surrounded by a coating $\Theta:=\Omega\setminus\overline{D}$ with a constant isotropic conductivity $\sigma_1$. }
We consider the conductivity problem
\beq\label{eqn:trans}
\begin{cases}
\ds\Delta\varphi_1 = 0 \quad&\mbox{in }\Theta, \\
\ds\nabla\cdot\bm{\sigma}_0\nabla\varphi_0=0\quad& \mbox{in } D,\\
\ds\varphi_0=\varphi_1\quad& \mbox{on }\p D,\\
\ds\left(\bm{\sigma}_0\nabla\varphi_0\right)\cdot \Bn=\sigma_1\nabla \varphi_1\cdot \Bn\quad&\mbox{on }\p D,\\
\ds\sigma_1\nabla \varphi_1\cdot \Bn=g\quad&\mbox{on }\p\Omega,
\end{cases}
\eeq  
where $g$ is a function which will be determined later, and $\Bn$ denotes the unit outward normal vector either to $\p\Om$ or to $\p D$. We further assume the uniformity condition
\beq\label{eqn:phi0}
\varphi_0(x_1,x_2)=-e_1 x_1 -e_2 x_2\quad\mbox{in }D
\eeq
for some real constants $e_1$ and $e_2$. 
The problem \eqnref{eqn:trans}-\eqnref{eqn:phi0} is over-determined, so that in general it has no solution for an arbitrary function $g$.
If a certain pair of domains $(\Om,D)$ admits a solution for some $e_1$, $e_2$ and $g$, then we call $D$ an {\it $E_{\Om}$-inclusion}. 
For later use, we denote 
\beq\label{ij_def}
\Be_0=(e_1,e_2)=-\nabla \varphi_0,\quad \Bj_0=(j_1,j_2)=\bm{\sigma}_0\Be_0
\eeq
for the uniform electric field and its associated current field inside $D$, respectively. We also set the complex numbers 
\beq\label{ij_def2}
e_0=e_1+\iu e_2,\quad j_0=j_1+\iu j_2.
\eeq

\begin{figure}[t]
 \centering
   \begin{subfigure}{0.35\textwidth}
      \centering
      \includegraphics[height=3.7cm]{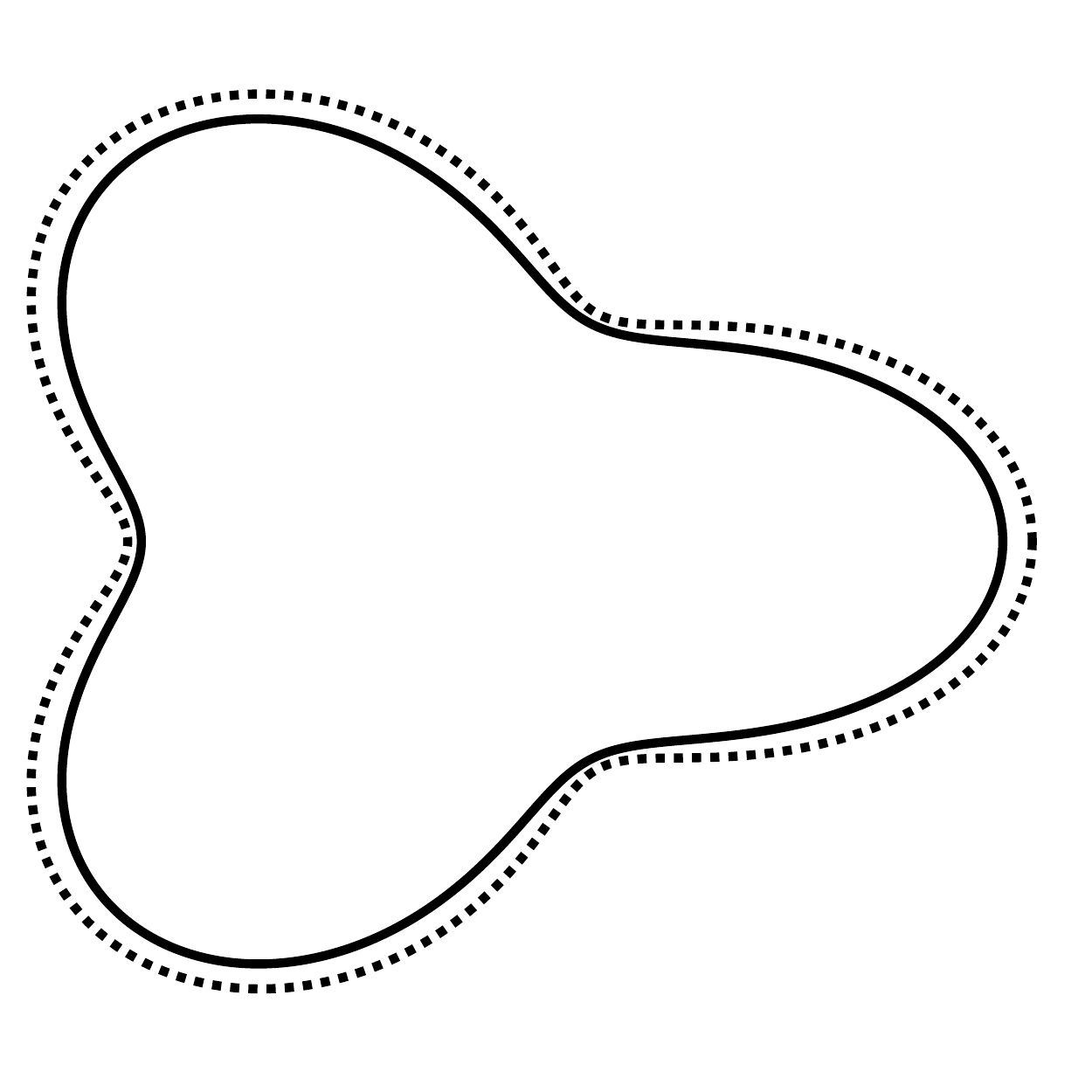}
      \vskip -1mm
     \caption{}
    \end{subfigure}
         \begin{subfigure}{0.35\textwidth}
      \centering
      \includegraphics[height=3.7cm]{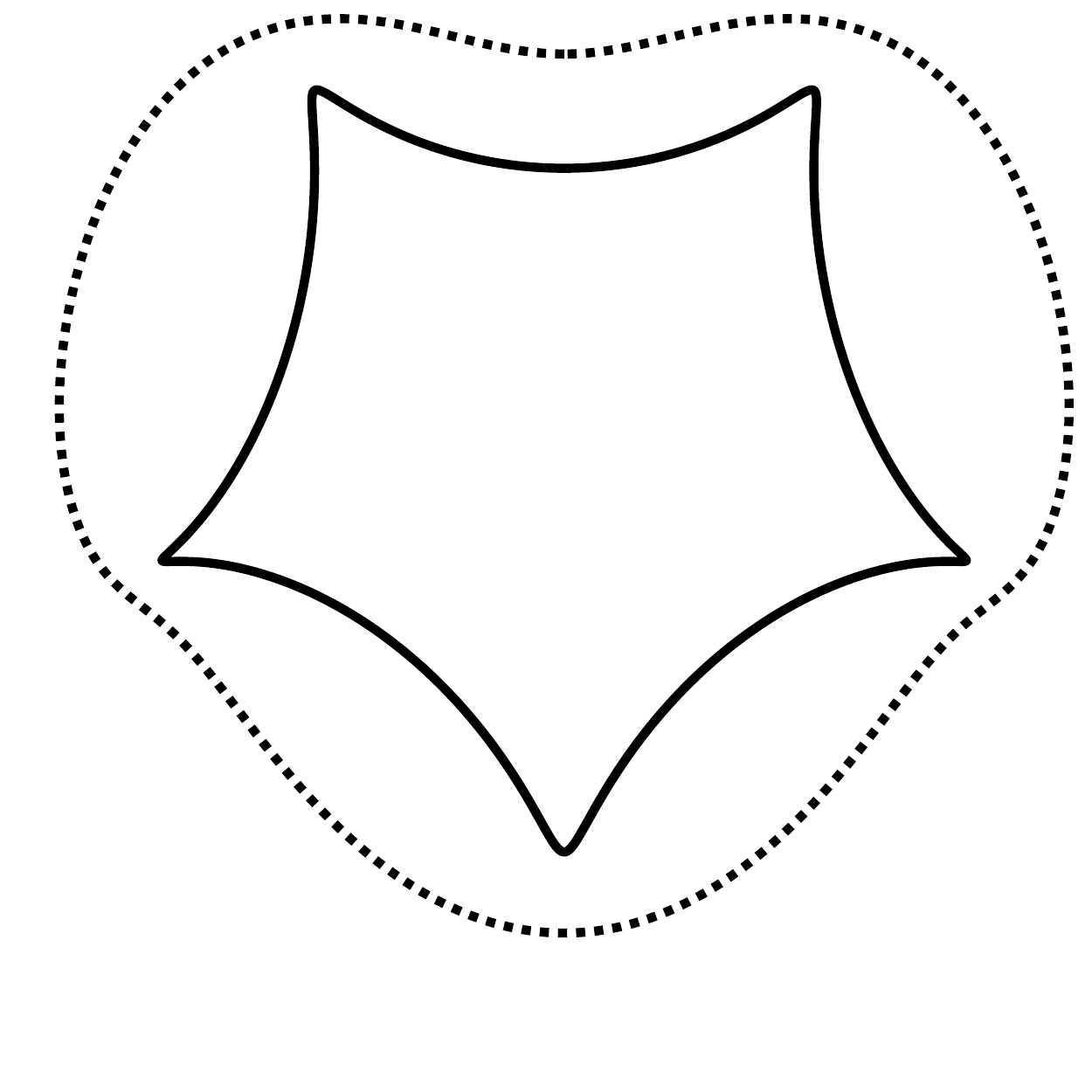}
      \vskip -1mm
      \caption{}
     \end{subfigure}
        \begin{subfigure}{0.35\textwidth}
      \centering
      \includegraphics[height=3.7cm]{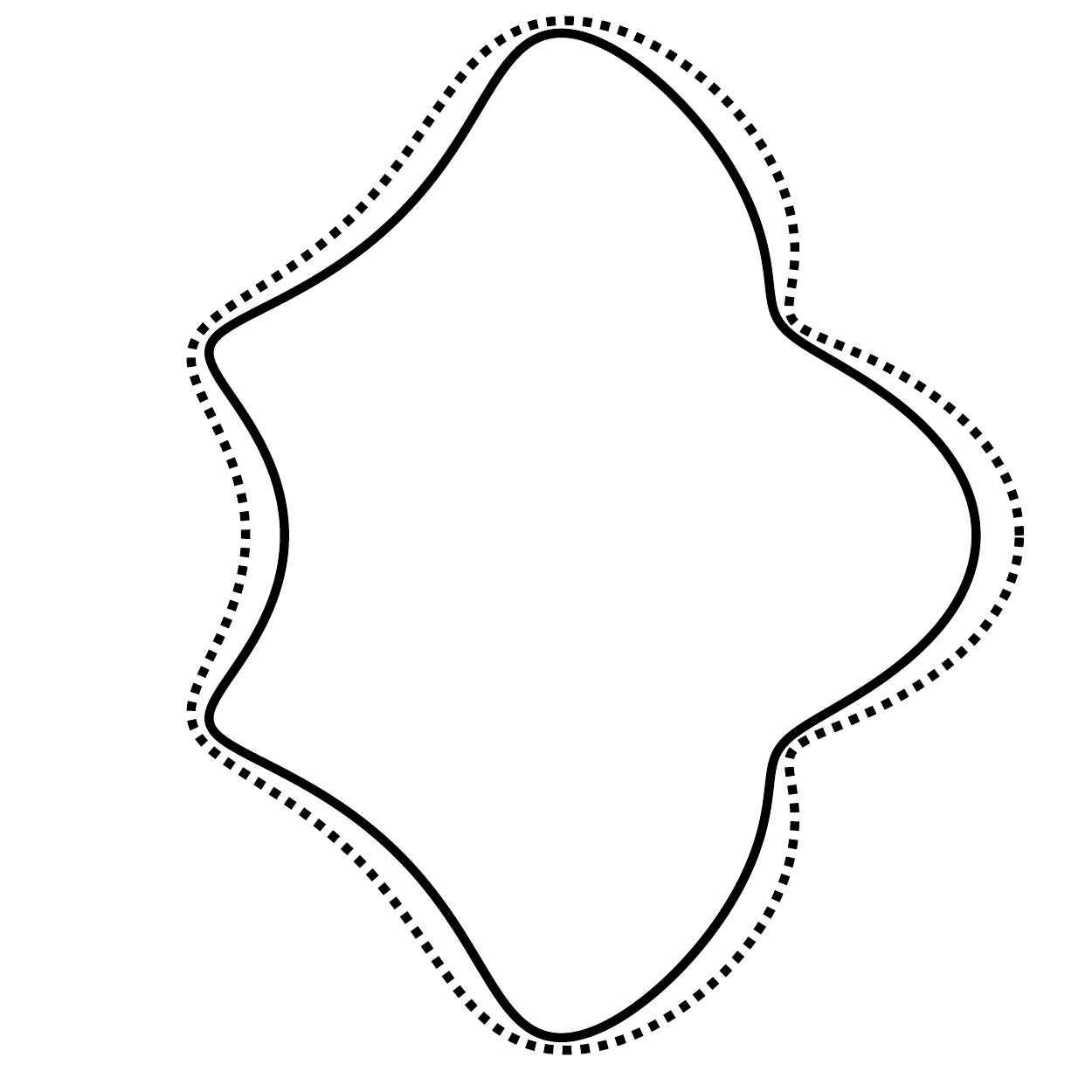}
      \vskip -1mm
      \caption{}
    \end{subfigure}
     \begin{subfigure}{0.35\textwidth}
      \centering
      \includegraphics[height=3.7cm]{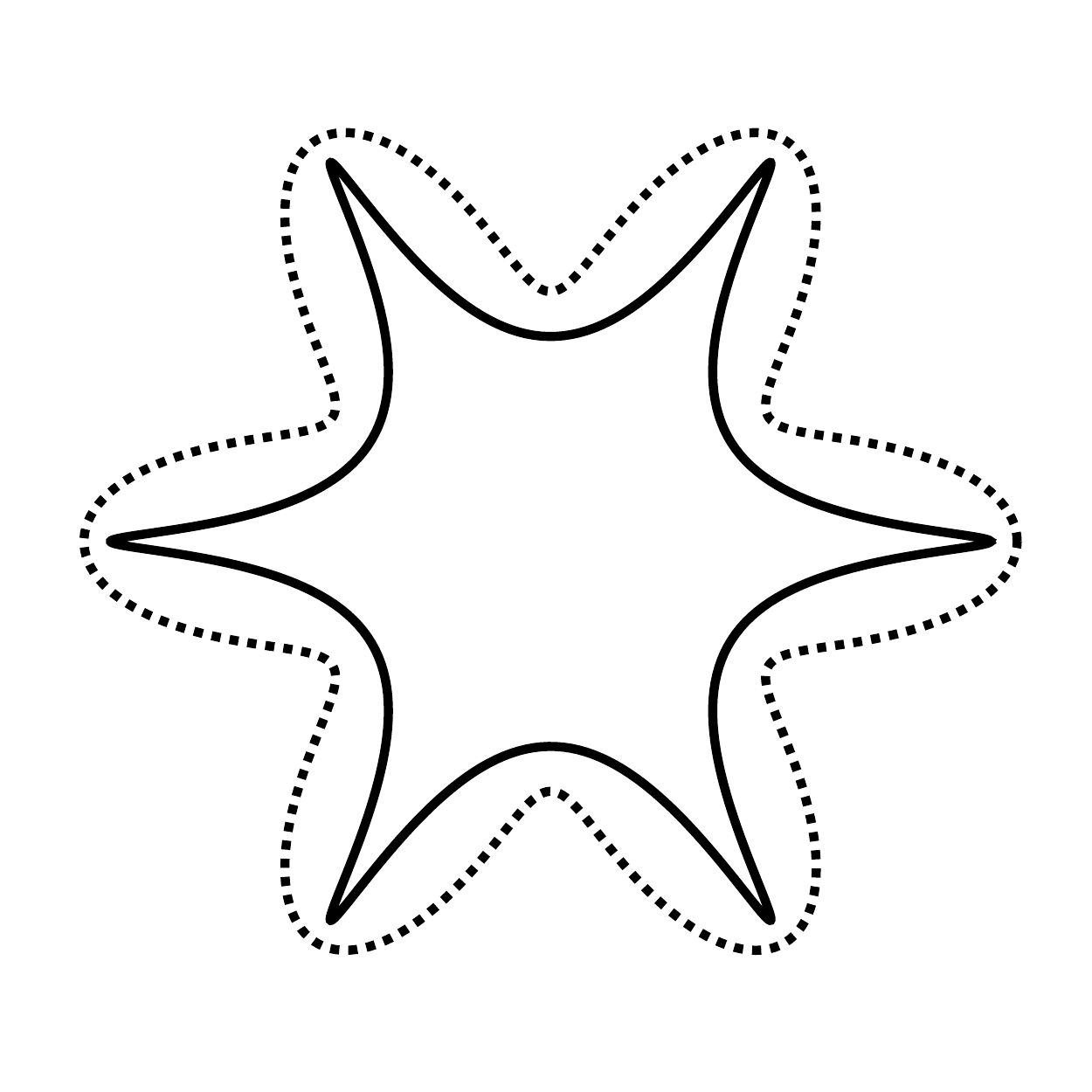}
      \vskip -1mm
      \caption{}
    \end{subfigure}
  
\caption{Various $E_\Omega$-inclusions (solid curve) with a possible boundary of $\Om$ (dotted curve) that can be generated from the formula \eqnref{Om:expression:2}.}\label{fig:various}
\end{figure}

In \cite{Bardsley:2017:CGB,Kang:2008:IPS}, $E_\Om$-inclusions were obtained by applying the hodograph transformation. Roughly speaking, in the hodograph transformation method one constructs the core $D$ by stretching a slit in the direction orthogonal to the slit.
In \cite{Bardsley:2017:CGB}, for example, a family of $E_\Om$-inclusions was constructed with $\p D$  parametrized by 
$$x_1=f\left(\pm\sqrt{\frac{1-x_2}{1+x_2}}\right),$$
where $f$ is a meromorphic function without a pole on the real axis. This formula gives rise to $E_\Om$-inclusions such that for all $(x_1,x_2)\in\p\Om$, except the extremal points $x_2=\pm 1$, each $x_2$ corresponds exactly to two $x_1$ values. 
In general, the boundary of $D$ obtained with the hodograph transformation method requires zero, one, or two intersecting points with any line that is orthogonal to the slit direction. 
In the present paper, however, we do not have such a restriction in the construction scheme, and it generates $E_\Om$-inclusions with an outer boundary of general shape as shown in Figure \ref{fig:various}.
Furthermore, we will show in section \ref{sec:EOm_prime} that any simply connected bounded domain is an $E_\Om$-inclusion for some $\Om$.

\subsection{Analytic function formulation}\label{sec:E_Om;formulation}
We can reformulate the over-determined problem \eqnref{eqn:trans}-\eqnref{eqn:phi0} in terms of complex analytic functions by using the fact that $\varphi_1$ is a two-dimensional harmonic function. 
In the following, we apply the existing complex potential approach \cite{Milton:2001:NCI}, where a free-boundary problem similar to \eqnref{eqn:trans} was solved to construct neutral coated inclusions.

As $\varphi_1$ has a mean-zero normal flux on $\p D$, it admits a single-valued harmonic conjugate, namely $\psi_1$, in $\Theta$ such that 
the complex function 
$$w(z)=\varphi_1(z)+\iu \psi_1(z)$$ 
is analytic. Hereafter, we identify $\Bx=(x_1,x_2)$ with $z=x_1+\iu x_2$. From the Cauchy-Riemann equations for complex analytic functions, we have
\beq\label{Psi_tau}
\pd{\psi_1}{\Bt}=\pd{\varphi_1}{\Bn}\quad\mbox{on }\p \Om,\p D,
\eeq
where $\Bt$ is the positively oriented unit tangent vector either on $\p \Om$ or on $\p D$. 
It is then straightforward to obtain from \eqnref{eqn:trans} that
\begin{align}\label{eqn:psiBC1}
\ds\pd{\psi_1}{\Bt}&=\ds\frac{1}{\sigma_1}g\quad\mbox{on }\p \Om.
\end{align}

The uniformity condition \eqnref{eqn:phi0} is essential for defining an $E_\Om$-inclusion.
Using \eqnref{eqn:phi0} together with \eqnref{Psi_tau} and the flux condition on $\p D$ in \eqnref{eqn:trans}, we can show that \beq\label{wz2}
w(z)=kz+h\bar{z}\quad\mbox{on }\p D
\eeq
with the complex constants $k$ and $h$ determined by the uniform electric field $\Be_0$ via the relations
\beq\label{gh}
k=-\frac{1}{2}\left(\overline{e_0}+\sigma_1^{-1}\overline{j_0}\right),\quad
h=\frac{1}{2}\left(-e_0+\sigma_1^{-1}j_0\right).
\eeq
Indeed, we have from \eqnref{eqn:trans}, \eqnref{ij_def} and \eqnref{Psi_tau} that
\begin{align*}
\pd{\psi_1}{\Bt}=\pd{\varphi_1}{\Bn}&=\frac{1}{\sigma_1} \left(\bm{\sigma}_0\nabla\varphi_0\right)\cdot\Bn
=\frac{1}{\sigma_1}\left(-\Bj_0\right)\cdot\Bn=\frac{1}{\sigma_1}(j_2,-j_1)\cdot\Bt\quad\mbox{on }\p D.
\end{align*}
This implies (the constant term is neglected)  
$$\psi_1(z)=\ds\frac{1}{\sigma_1}\left(j_2x_1-j_1x_2\right)\quad \mbox{on }\p D.$$
Hence, we can show \eqnref{wz2}-\eqnref{gh} by using \eqnref{eqn:phi0}.
It is worth mentioning that $h$ and $k$ can inversely determine $e_0$ and $j_0$ as
\beq\notag
e_0=-h-\overline{k},\quad j_0=\sigma_1(h-\overline{k}).
\eeq

Because $\Theta$ is a doubly connected domain, it is conformally equivalent to an annulus $\left\{p: r<|p|<R\right\}$ for some $0<r<R$. 
In other words, there is a conformal mapping, namely $z(p)$, from the annulus onto $\Theta$.
As $z(p)$ is analytic on the annulus that is centered at zero, it admits a Laurent series expansion
\beq\label{series:z2}
z(p)=\sum_{n=-\infty}^\infty a_n p^n,\quad r<|p|<R,
\eeq
with some complex coefficients $a_n$. 
The composition $w(p):=w(z(p))$ is also analytic in the annulus and, hence, admits a Laurent series expansion
\beq\label{series:w2}
w(p)=\sum_{n=-\infty}^{\infty}b_n p^n,\quad r<|p|<R,
\eeq
where the coefficients $b_n$ should be given for $w$ to satisfy the boundary constraint \eqnref{wz2}. 
The condition \eqnref{wz2} is equivalent to
\beq\label{eqn:bn}
    b_n = k a_n + hr^{-2n}\overline{a_{-n}}\quad \mbox{for all }n\in\ZZ,
\eeq
and it uniquely determines $b_n$ for given $a_n$, $k$ and $h$.

We can now construct $E_\Om$-inclusions by specifying the coefficients $a_n$, which should be chosen such that the resulting Laurent series $z(p)$ converges to a conformal mapping from an annulus to a doubly connected domain, and such that the series  \eqnref{series:w2} for $w(p)$ with coefficients $b_n$ given by \eqnref{eqn:bn} converges in this annulus.
We set the pair of domains $(\Om, D)$ as
\beq\label{Om:expression:2}
\p \Om = \{z(p):|p|=R\},\quad \p D =\{z(p):|p|=r\}
\eeq
and define $w(p)$ by using the formula \eqnref{eqn:bn}, with $h$ and $k$ given by \eqnref{gh}, for a given uniform field $\varphi_0$.
Given that the resulting series function $w(p)$ also converges to an analytic function in the annulus, the function
\beq\label{varphi1:w}
\varphi_1(z)=\Re\{w(z)\}, \quad z\in\Om\setminus\overline{D},
\eeq
satisfies the over-determined problem \eqnref{eqn:trans}-\eqnref{eqn:phi0} with 
\beq\label{g:expression}
g =\sigma_1 \pd{\Im\{w\}}{\Bt}\quad\mbox{on }\p\Om. 
\eeq
The series $w(p)$ converges if the two series $z(p)=\sum a_n p^n$ and 
$\sum r^{-2n}\overline{a_{-n}}\, p^n$ 
converge.
Hence, the convergence of $w(p)$ is independent of the direction of the uniform field $\varphi_0$. In other words, the constructed $E_\Om$-inclusions admit an arbitrary uniform field inside the core $D$, where the assignment of the flux function $g$ depends on the direction of the uniform field.

\subsection{$E_\Om$-inclusions with a core of arbitrary shape}\label{sec:EOm_prime}
\begin{figure}[htp!]
 \centering
\includegraphics[ height=4cm, width=4cm, trim={0 0 0 0},clip]{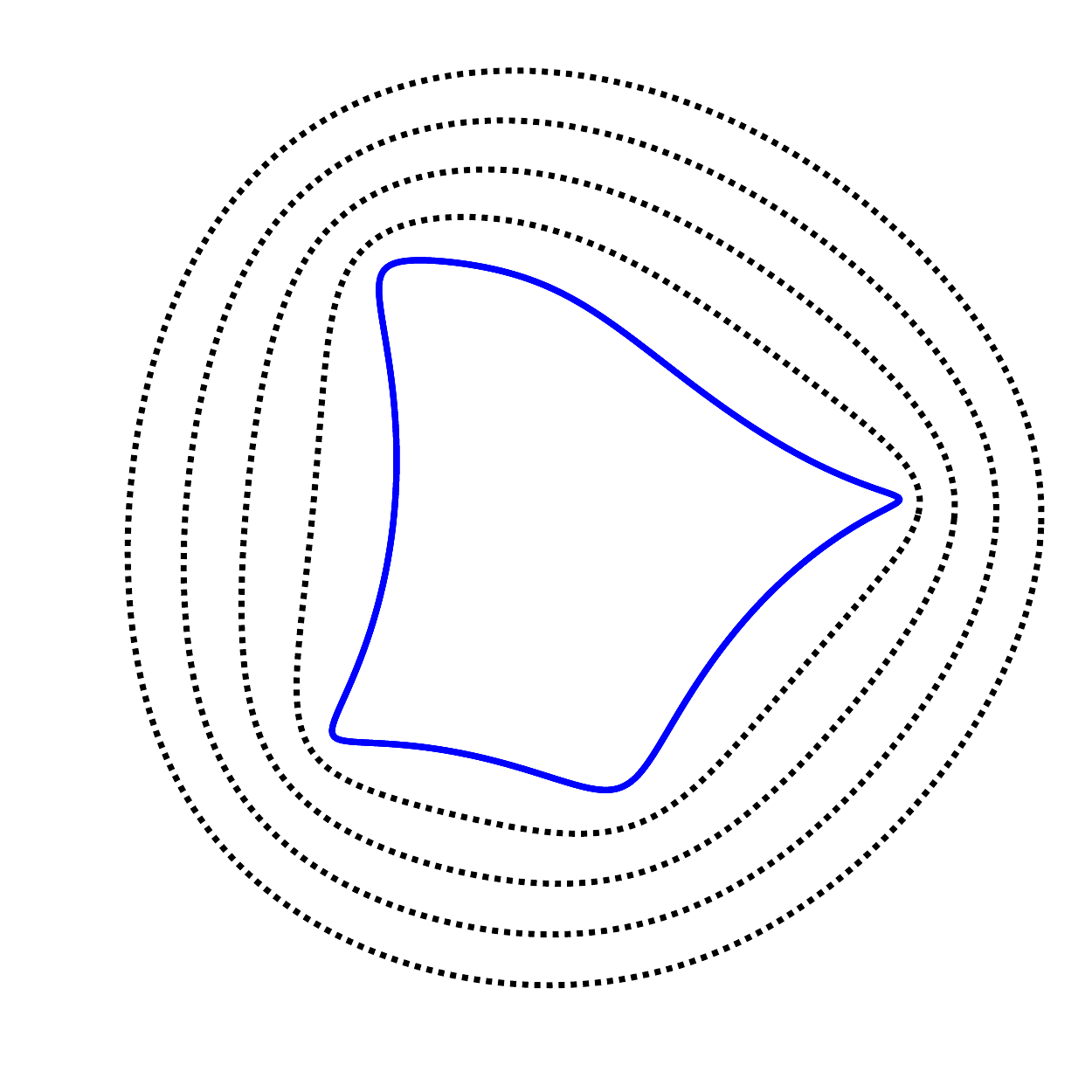}
\caption{
An $E_\Om$-inclusion $D$ (blue solid curve), with the black dotted curves being possible boundaries of $\Om$. Here, $a_n$ has only a finite number of non-zero entries and, hence, $w(p)$ given by \eqnref{E_Om:general:w} is convergent; thus $\Omega$ can be taken to be any region that encloses the inclusion.}\label{fig:EOmOmprime}
\end{figure}

The proposed construction scheme enables us to find an $E_\Om$-inclusion with a core of arbitrary analytic shape. 
Let $D$ be an arbitrary simply connected domain. Then consider the conformal mapping of the form 
$$z(p)=p+\sum_{n=-\infty}^{0} a_n p^{n}$$
that takes the exterior of the unit disk centered at the origin onto the exterior of $D$ in a bijective fashion. 
{We give an additional regularity assumption on $D$ that $z(p)$ is univalent analytic outside a smaller disk $\left\{p:|p|>\rho^*\right\}$ for some $\rho^*<1$,}
the associated Laurent series (with $r=1$) for the potential
\beq\label{E_Om:general:w}
w(p)=kp+h/p+\sum_{n=-\infty}^{0} ka_n p^n +\sum_{n=0}^\infty h\overline{a_{-n}}p^n
\eeq
is analytic in $\left\{p:1/{\rho^*}>|p|>\rho^*\right\}$. The domain of analyticity of $w(p)$ is almost certainly larger than this, and given any Jordan
curve $\gamma$ enclosing the unit disk such that $w(p)$ is analytic in the annular region between $\gamma$ and the unit disk, we see that $D$ is an $E_\Om$-inclusion with
the boundary of $\Om$ given by
$$\p\Om:=\{z(p): p\in \gamma\}.$$
Fig.\;\ref{fig:EOmOmprime} shows several possible boundaries of $\Om$. 

It is worth remarking that we can interpret \eqnref{eqn:trans}-\eqnref{eqn:phi0}
as a Cauchy problem: for given $\varphi_0$ of the form \eqnref{eqn:phi0}, find $\varphi_1$ such that 
\beq\label{eqn:Cauchy}
\begin{cases}
\ds\Delta\varphi_1 = 0 \quad&\mbox{in }\Theta, \\
\ds\varphi_1=\varphi_0\quad& \mbox{on }\p D,\\
\ds \pd{\varphi_1}{\Bn} =\frac{1}{\sigma_1} \left(\bm{\sigma}_0\nabla\varphi_0\right)\cdot \Bn\quad&\mbox{on }\p D.
\end{cases}
\eeq 
We then assign $g$ in \eqnref{eqn:trans} in terms of the solution $\varphi_1$ to \eqnref{eqn:Cauchy}. 
The well-known Cauchy-Kovalevskaya theorem ensures the local solvability of the general Cauchy problem for partial differential equations.
Cauchy problems for elliptic problems have been extensively studied, for example, see \cite{Alessandrini:2009:SCP, Kozlov:1991:IMS, Nguyen:2015:CAL, Nguyen:2017:DIT, Vabishchevich:1978:SCP}.
The analysis in this subsection enables us to explicitly find the possible regions $\Theta$ beyond the vicinity of $\p D$ in which the Cauchy condition extends to a solution for the uniform field case.

\subsection{Numerical Examples}

Fig.\;\ref{fig:various} shows $E_\Omega$-inclusions of various shapes. We emphasize that 
the star-shaped domain $D$ in Fig.\;\ref{fig:various}(d) cannot be achieved by applying the hodograph transformation (which is used in \cite{Bardsley:2017:CGB,Kang:2008:IPS}). 
As previously explained, for the $E_\Om$-inclusions constructed with the hodograph transformation, the outer boundary of $D$ requires zero, one, or two intersecting points with any line that is orthogonal to the slit direction. 

\begin{figure}[htp!]
 \centering
   \begin{subfigure}{0.35\textwidth}
      \centering
      \includegraphics[height=3.5cm]{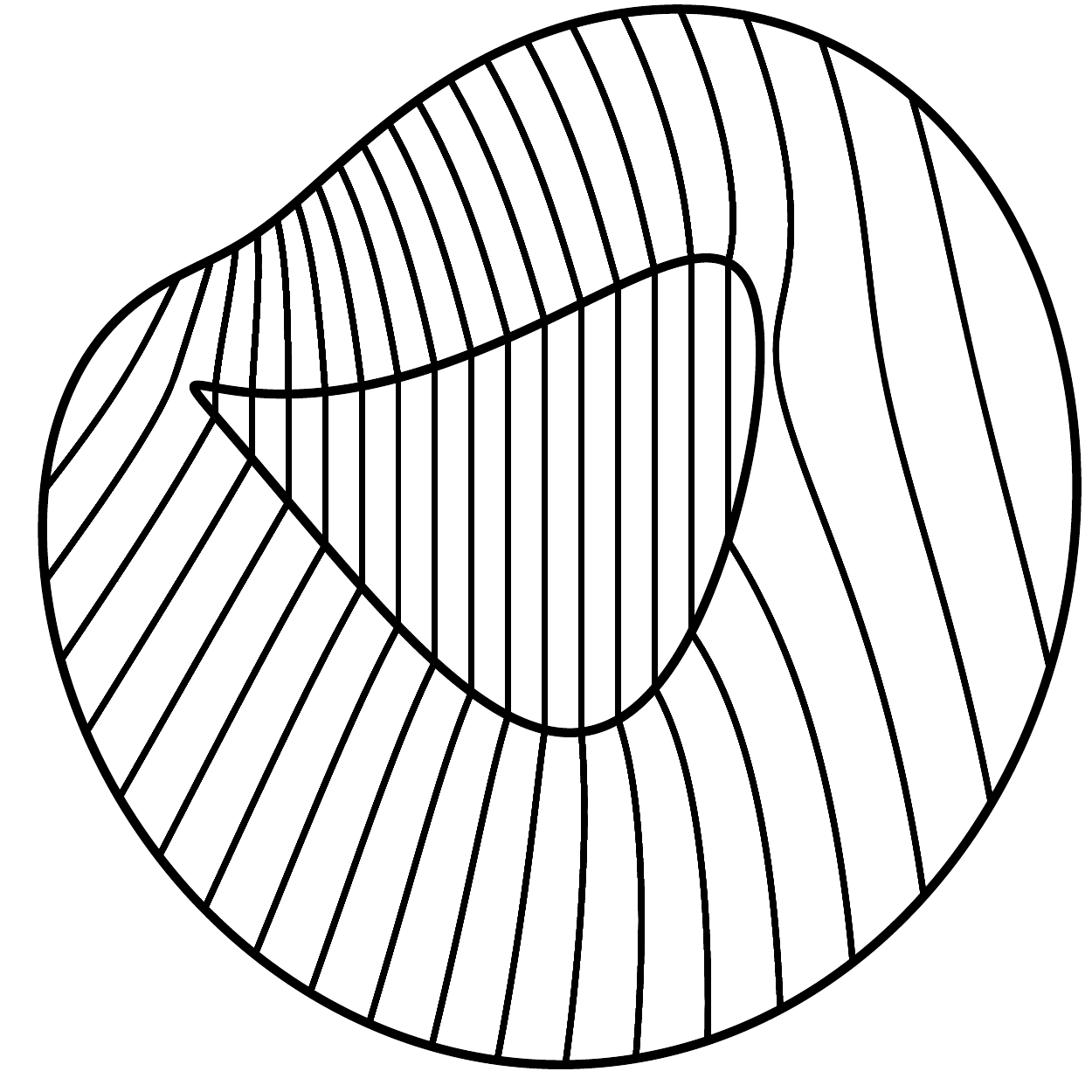}
      \caption{}
    \end{subfigure}
        \begin{subfigure}{0.35\textwidth}
      \centering
      \includegraphics[height=3.5cm]{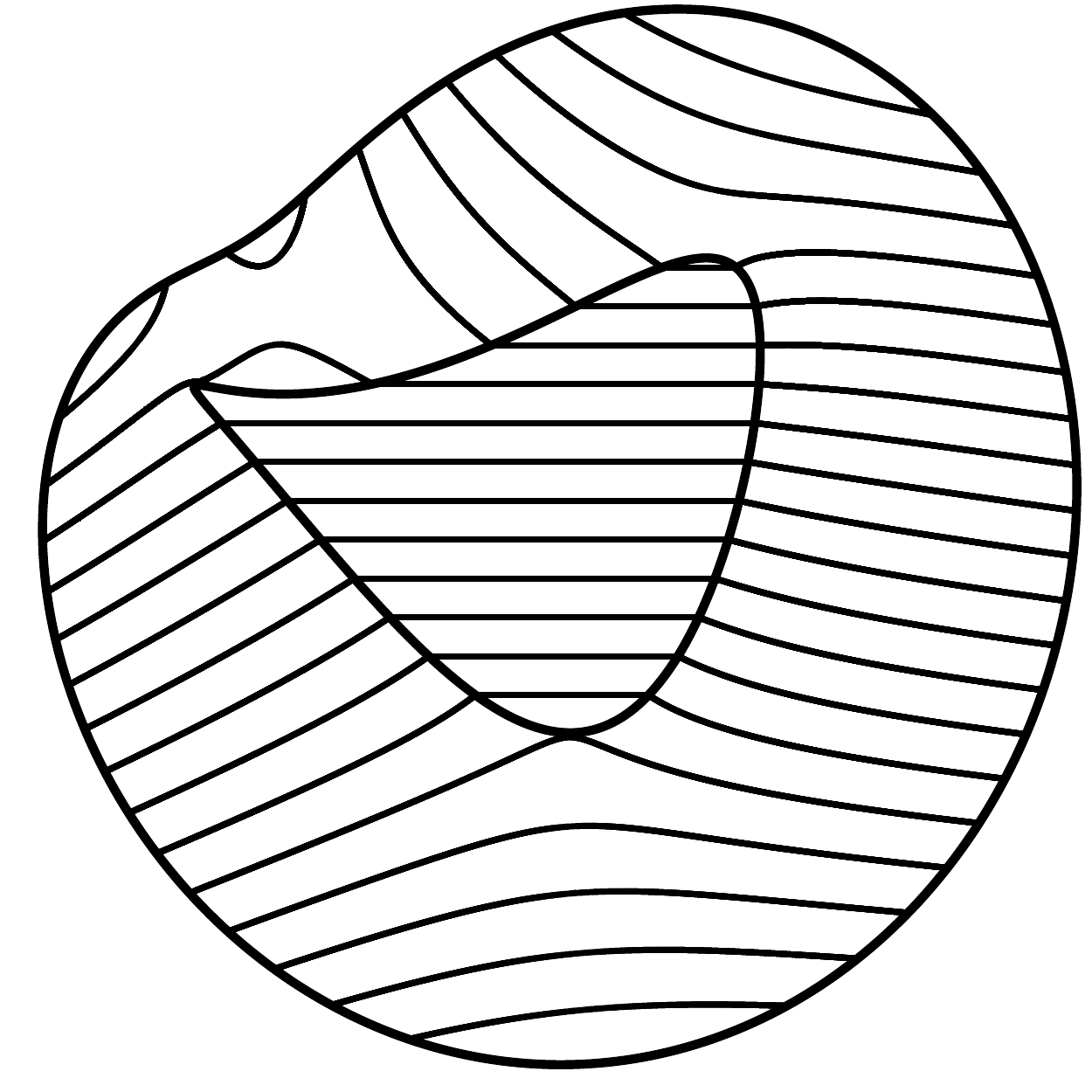}
      \caption{}
    \end{subfigure}
     \begin{subfigure}{0.35\textwidth}
      \centering
      \includegraphics[height=3.5cm]{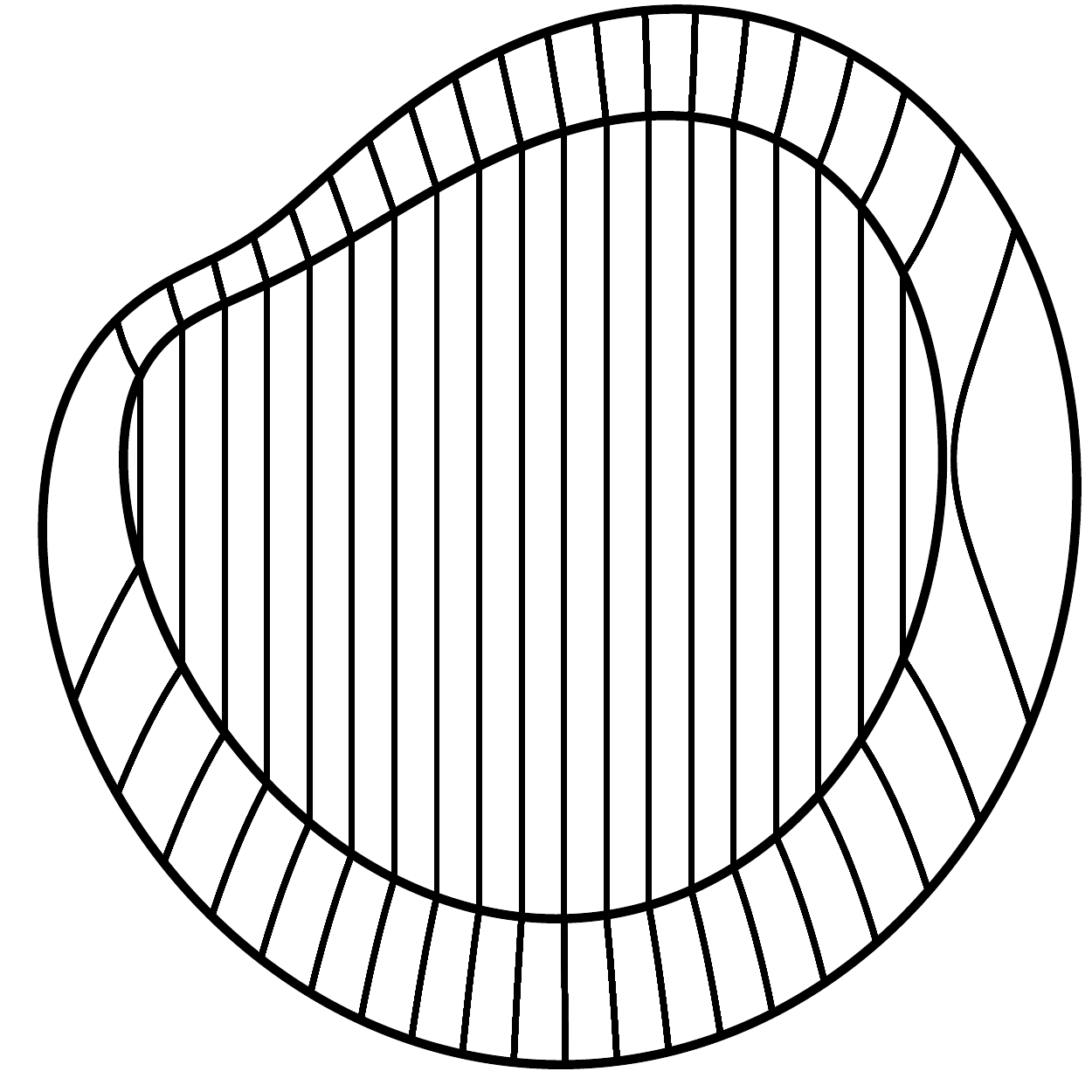}
      \caption{}
    \end{subfigure}
        \begin{subfigure}{0.35\textwidth}
      \centering
      \includegraphics[height=3.5cm]{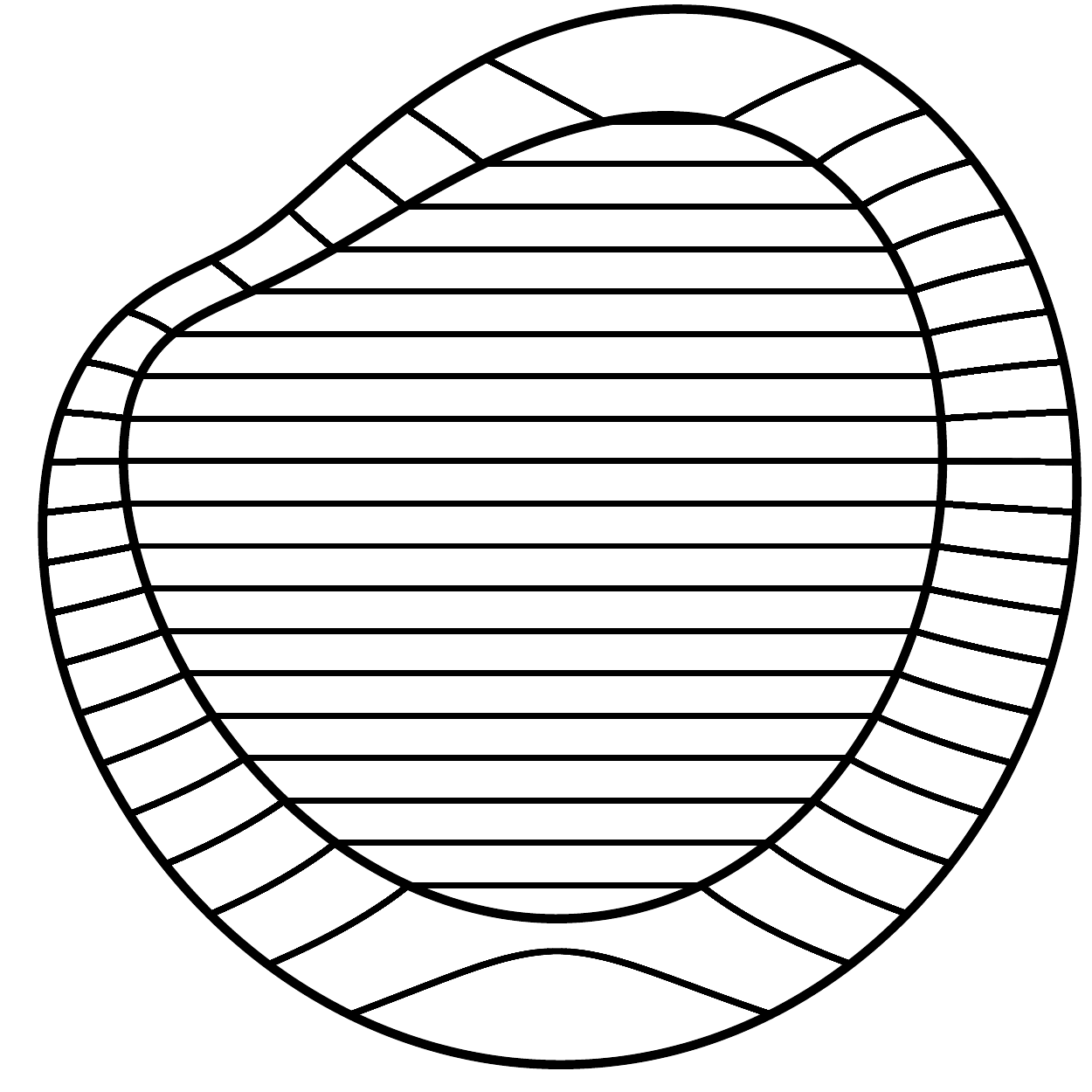}
      \caption{}
    \end{subfigure}   
       \vskip -1mm
\caption{$E_\Om$-inclusions. The figures illustrate $\p D$, $\p\Om$ and the current flow (equipotential lines of $\varphi_1$). The pairs $(\Omega,D)$ and the boundary flux $g$ are given by \eqnref{Om:expression:2} and \eqnref{g:expression}. The nonzero geometric parameters are $a_{-2}=-0.03 - 0.03\iu,\ a_{-1}= 0.06+0.06\iu,\ a_1 = 1-\iu$, and $a_2=0.3-0.3\iu$. We fix $\sigma_1/\sigma_0=4$, $R=1$ and set $r$ as follows: (a), (b) $r=0.5$; (c), (d) $r=0.8$.}\label{fig:E_Om}
 \vskip .3cm
 \centering
   \begin{subfigure}{0.4\textwidth}
      \hskip -2.5mm
      \includegraphics[height=4.4cm, width=5.1cm]{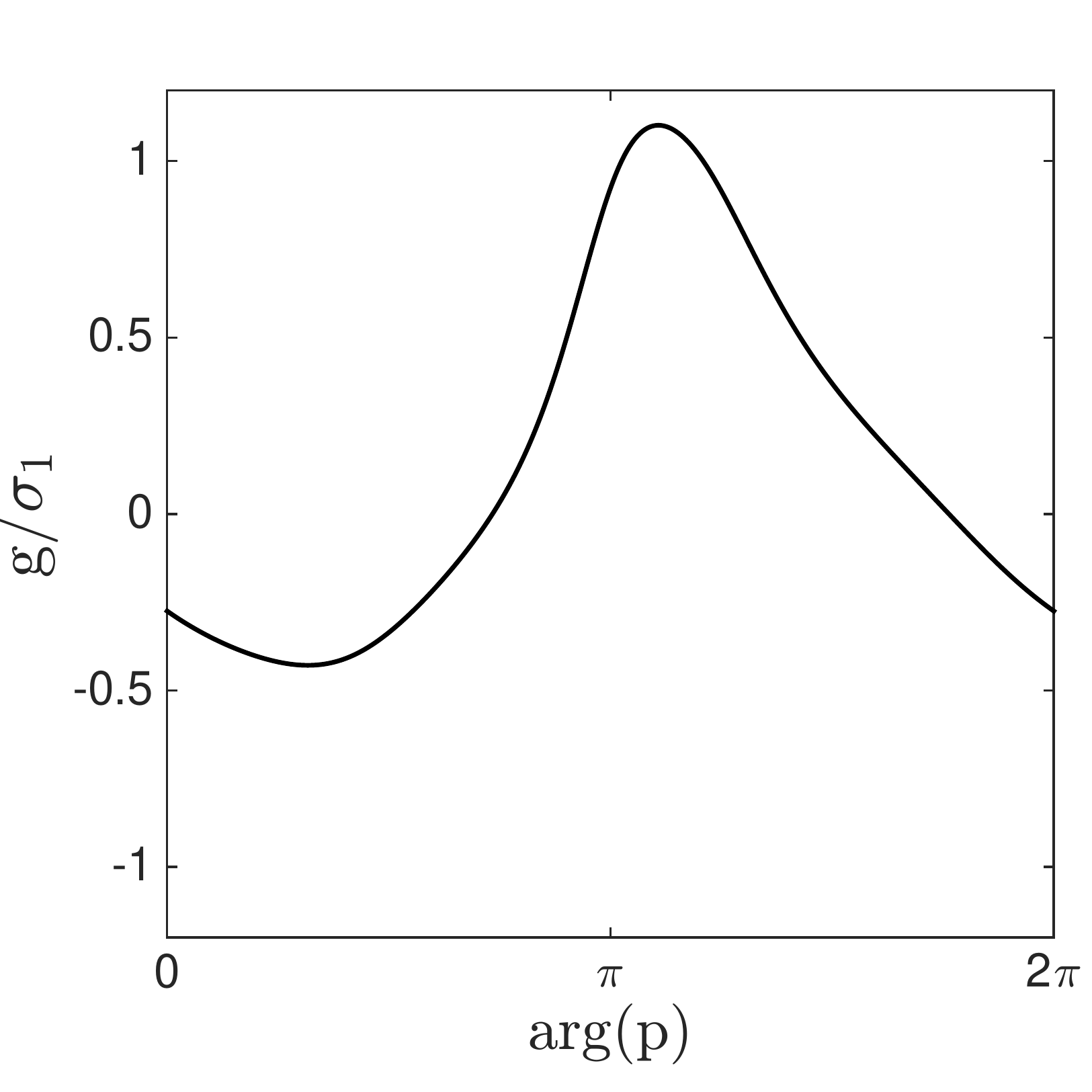}
      \caption{}
    \end{subfigure}
        \begin{subfigure}{0.4\textwidth}
     \hskip -2.5mm
      \includegraphics[height=4.4cm, width=5.1cm]{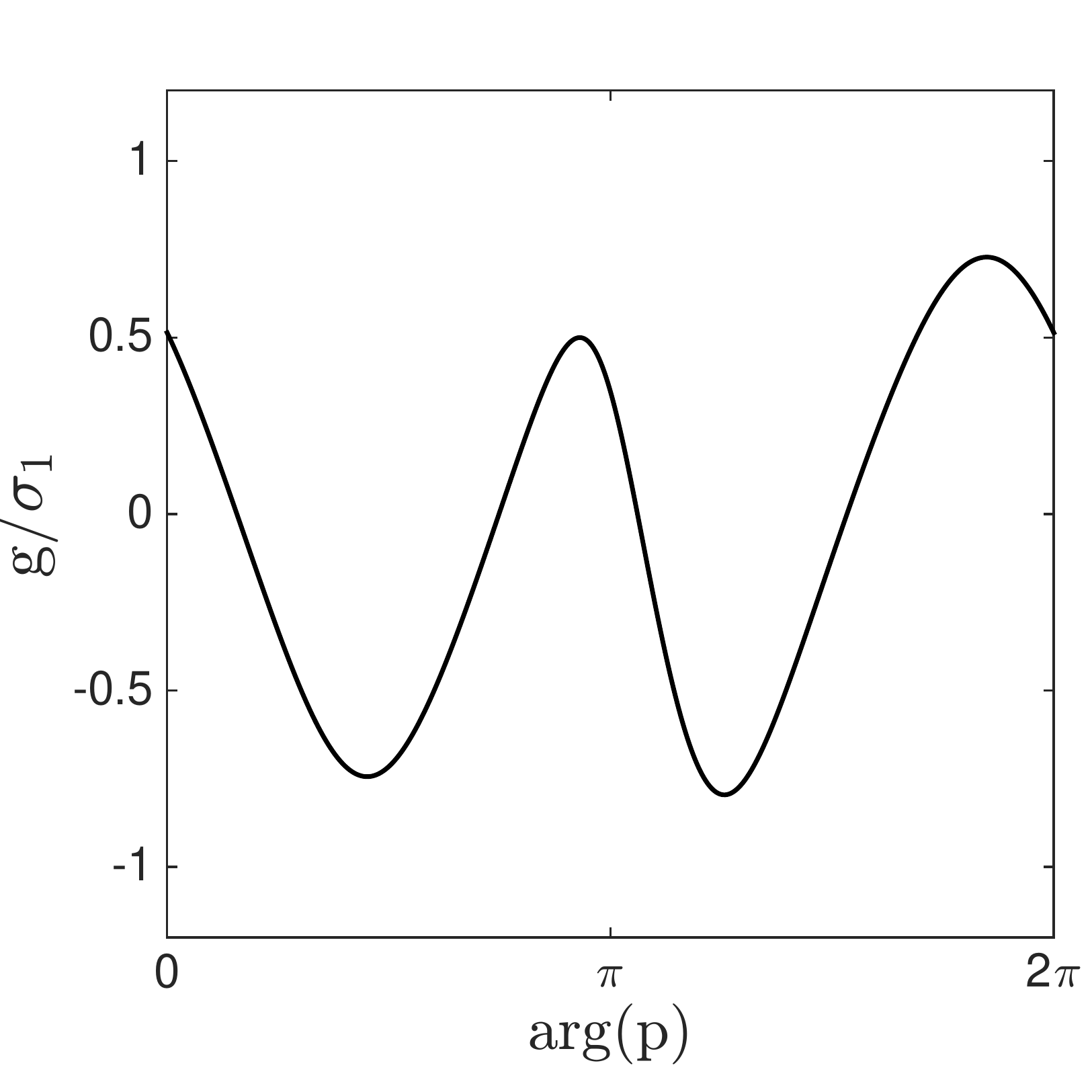}
      \caption{}
    \end{subfigure}
    \vskip -.3cm
     \begin{subfigure}{0.4\textwidth}
      \hskip -2.5mm
      \includegraphics[height=4.4cm, width=5.1cm]{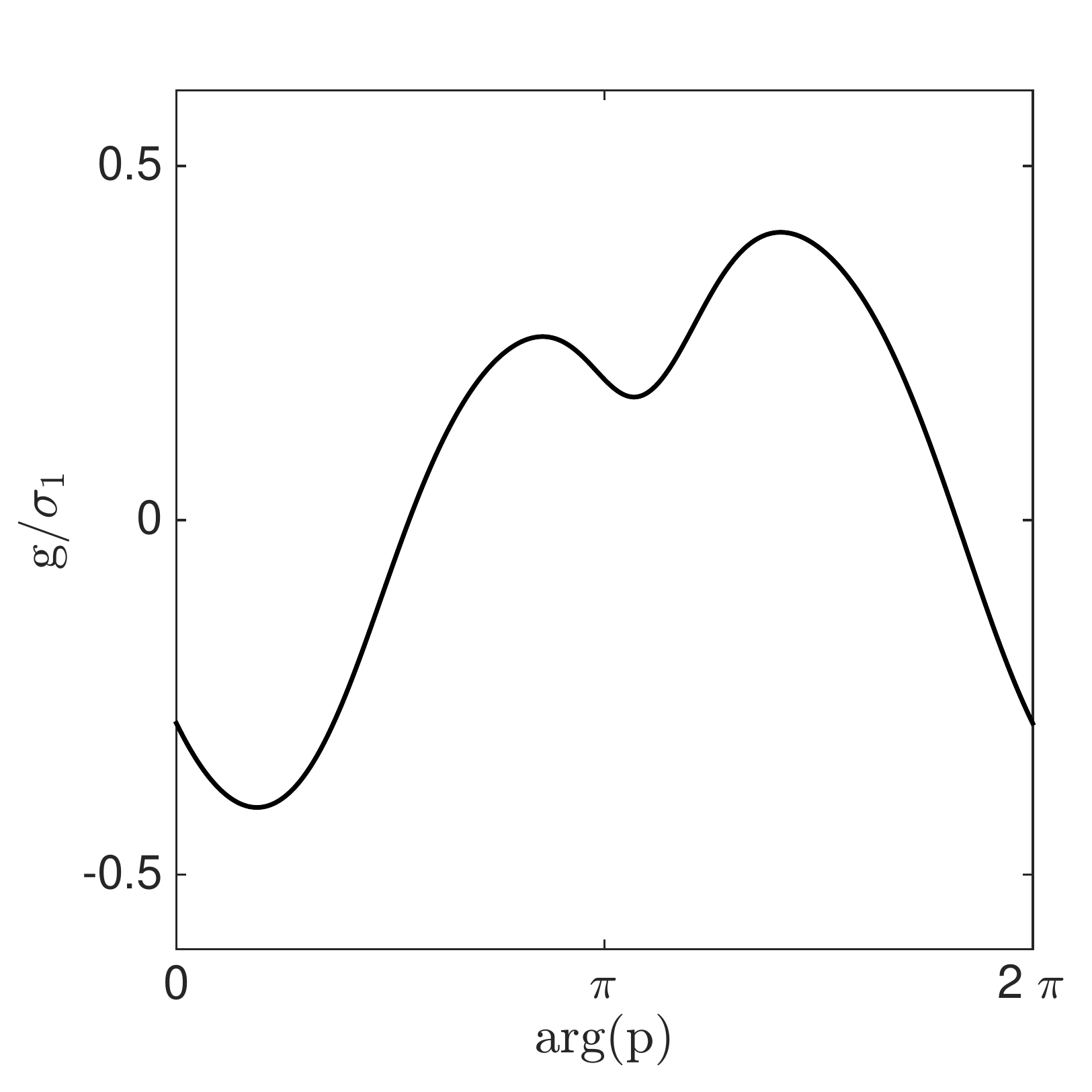}
      \caption{}
    \end{subfigure}
    \hskip .2cm
        \begin{subfigure}{0.4\textwidth}
      \hskip -2.5mm
      \includegraphics[height=4.4cm, width=5.1cm]{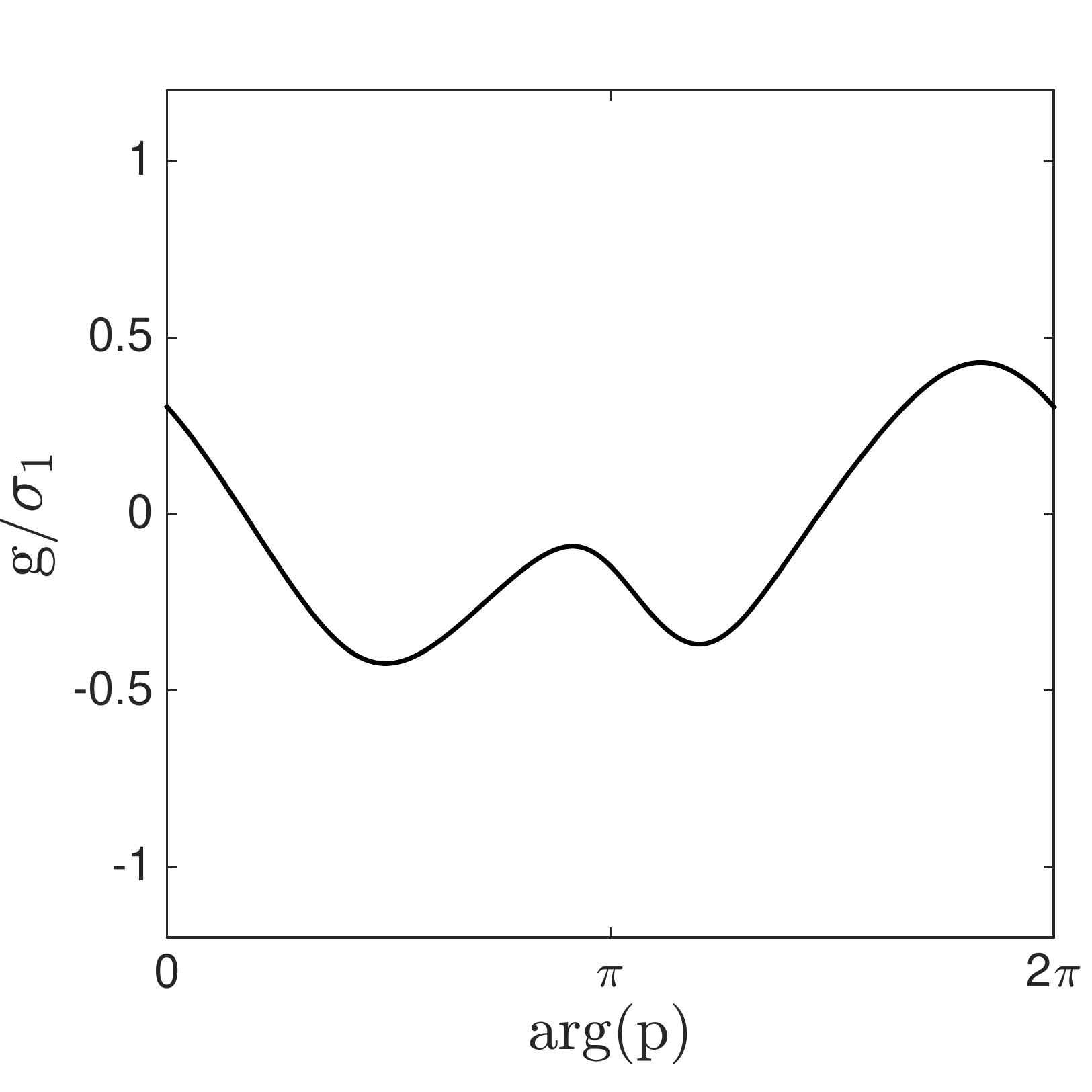}
      \caption{}
    \end{subfigure}
   \vskip -3mm
\caption{Flux function $g$ corresponding to the examples in Fig.\;\ref{fig:E_Om}.}\label{fig:E_Om_flux}\end{figure}

\begin{figure}[htp!]
 \centering
   \begin{subfigure}{0.35\textwidth}
      \centering
      \includegraphics[height=3.5cm]{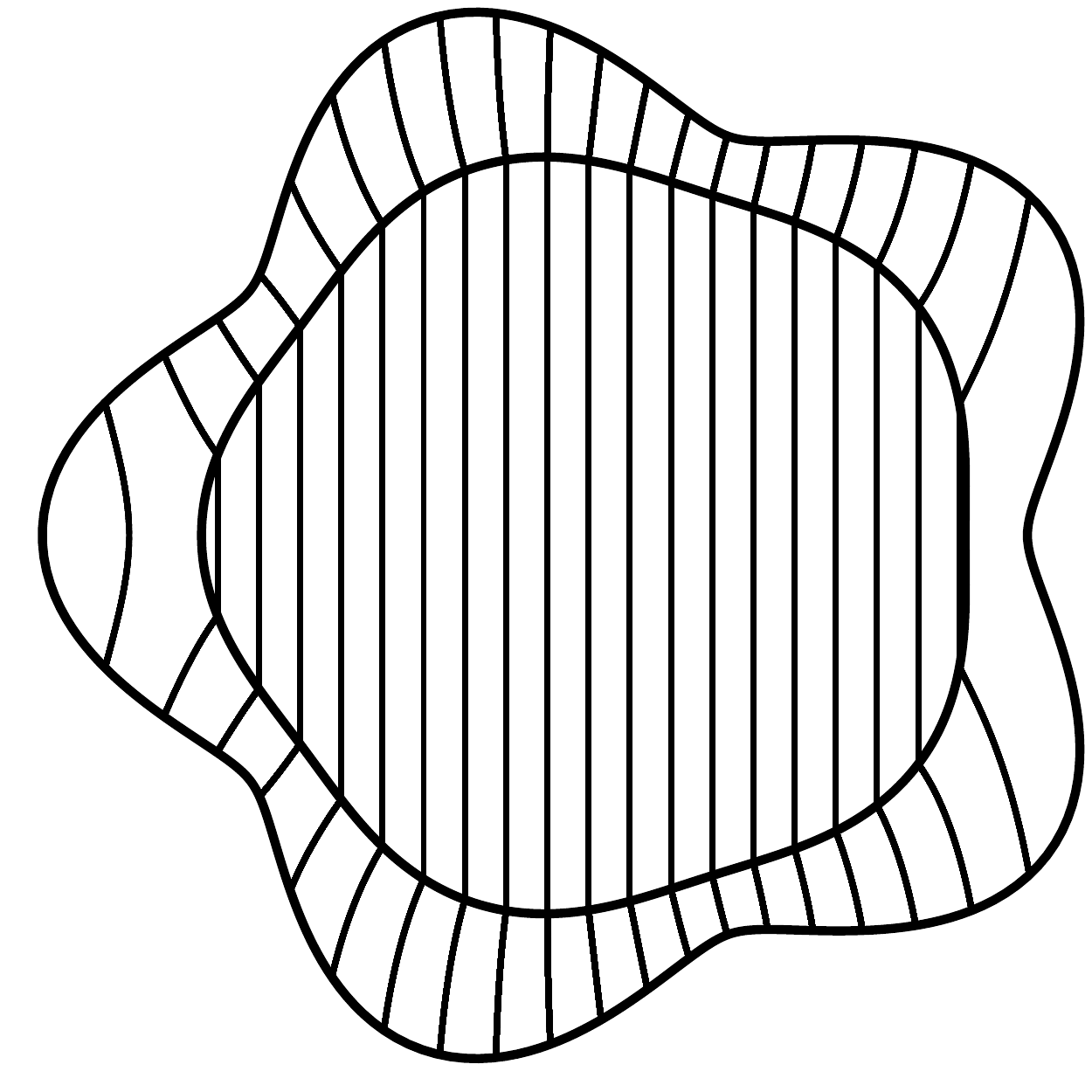}
      \caption{}
    \end{subfigure}
        \begin{subfigure}{0.35\textwidth}
      \centering
      \includegraphics[height=3.5cm]{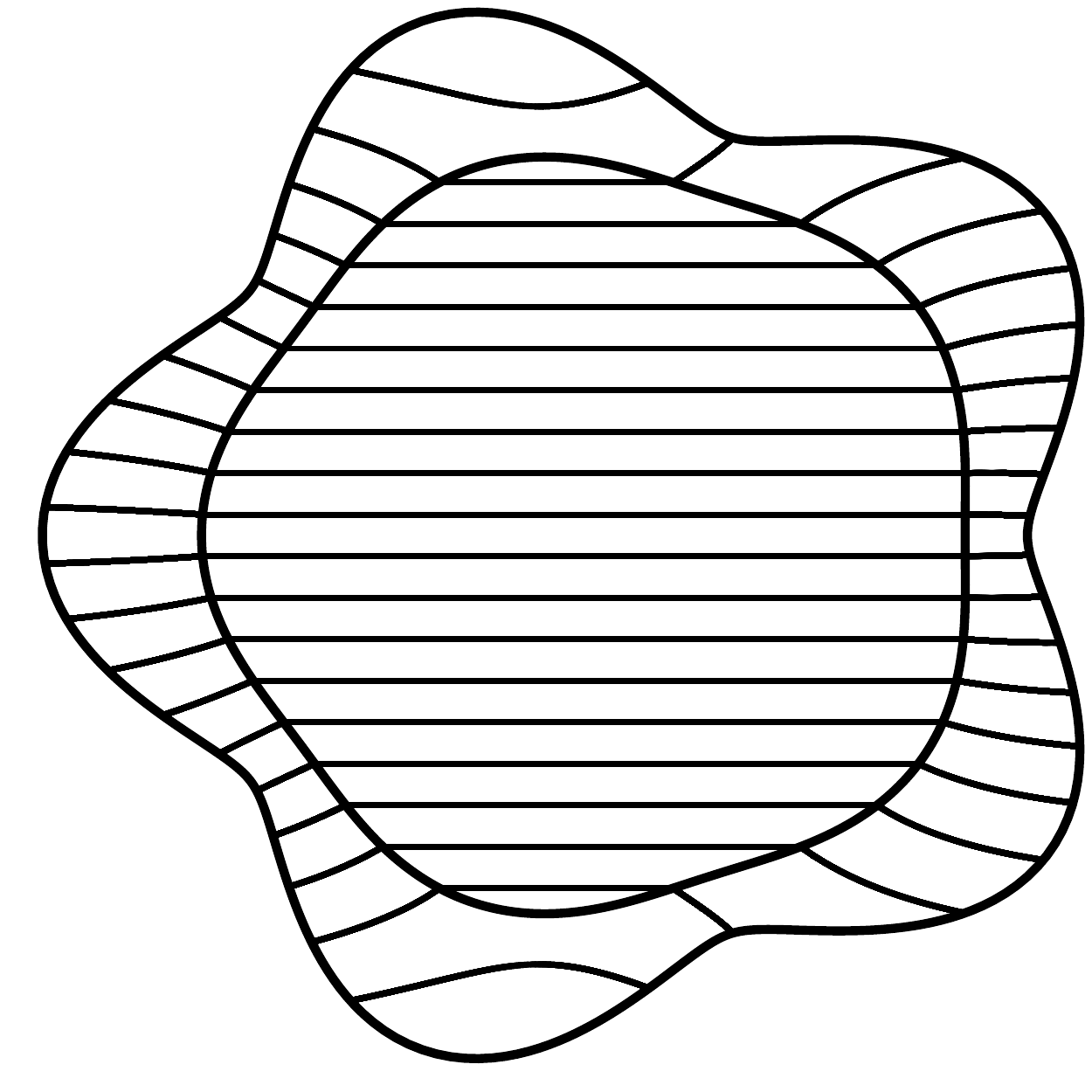}
      \caption{}
    \end{subfigure}
     \begin{subfigure}{0.35\textwidth}
      \centering
      \includegraphics[height=3.5cm]{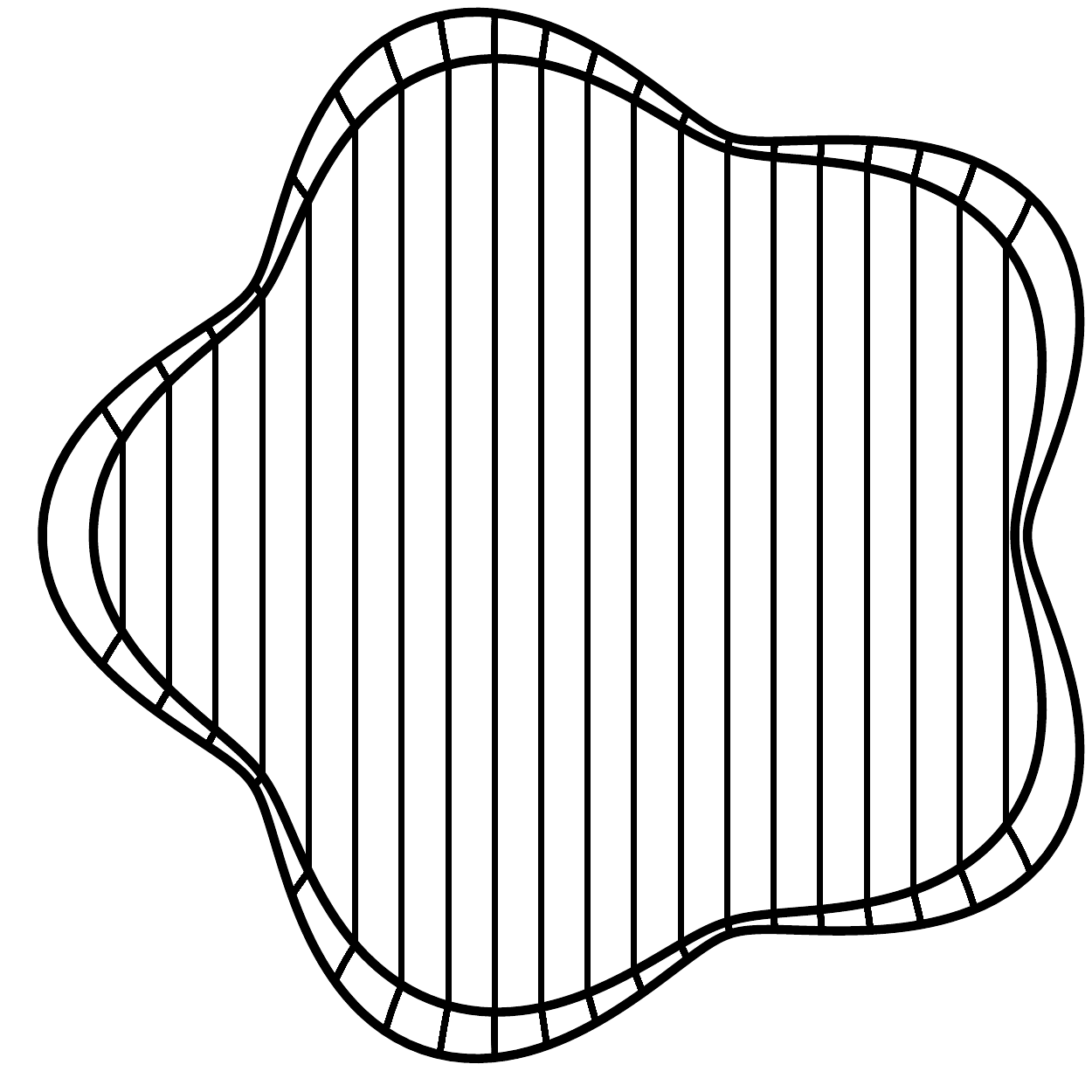}
      \caption{}
    \end{subfigure}
        \begin{subfigure}{0.35\textwidth}
      \centering
      \includegraphics[height=3.5cm]{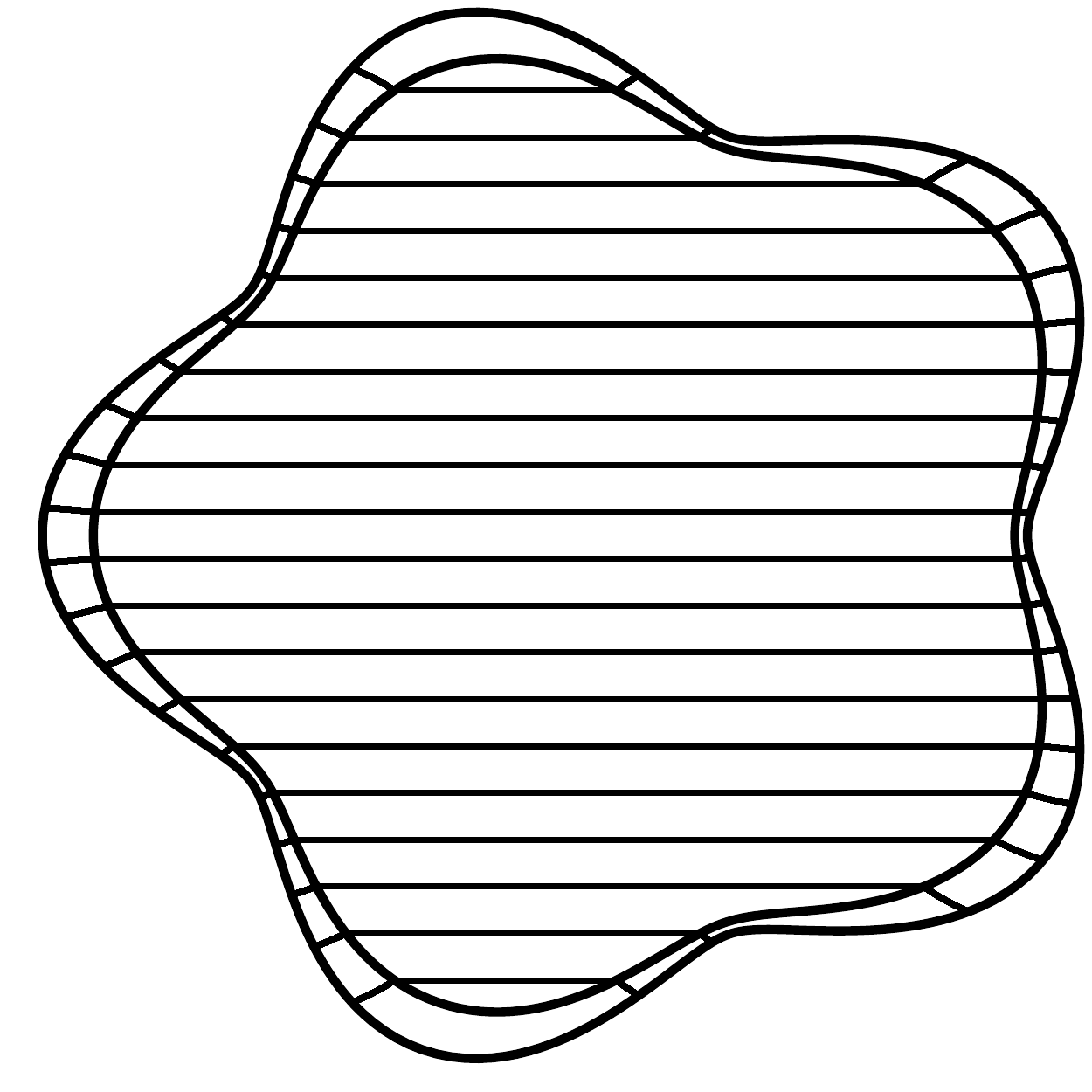}
      \caption{}
    \end{subfigure}

        \vskip -1mm
\caption{$E_\Om$-inclusions. The figures illustrate $\p D$, $\p\Om$ and the current flow (equipotential lines of $\varphi_1$). The pairs $(\Omega,D)$ and the boundary flux $g$ are given by \eqnref{Om:expression:2} and \eqnref{g:expression} with $a_n$ in \eqnref{def:a_n}. We fix $\sigma_1/\sigma_0=4$, $R=1$ and set $r$ as follows: (a), (b) $r=0.8$; (c), (d) {$r=0.95$.}}\label{fig:E_Om_star}
 \vskip .3cm
 \centering
   \begin{subfigure}{0.4\textwidth}
      \hskip -2.5mm
      \includegraphics[height=4.4cm, width=5.1cm]{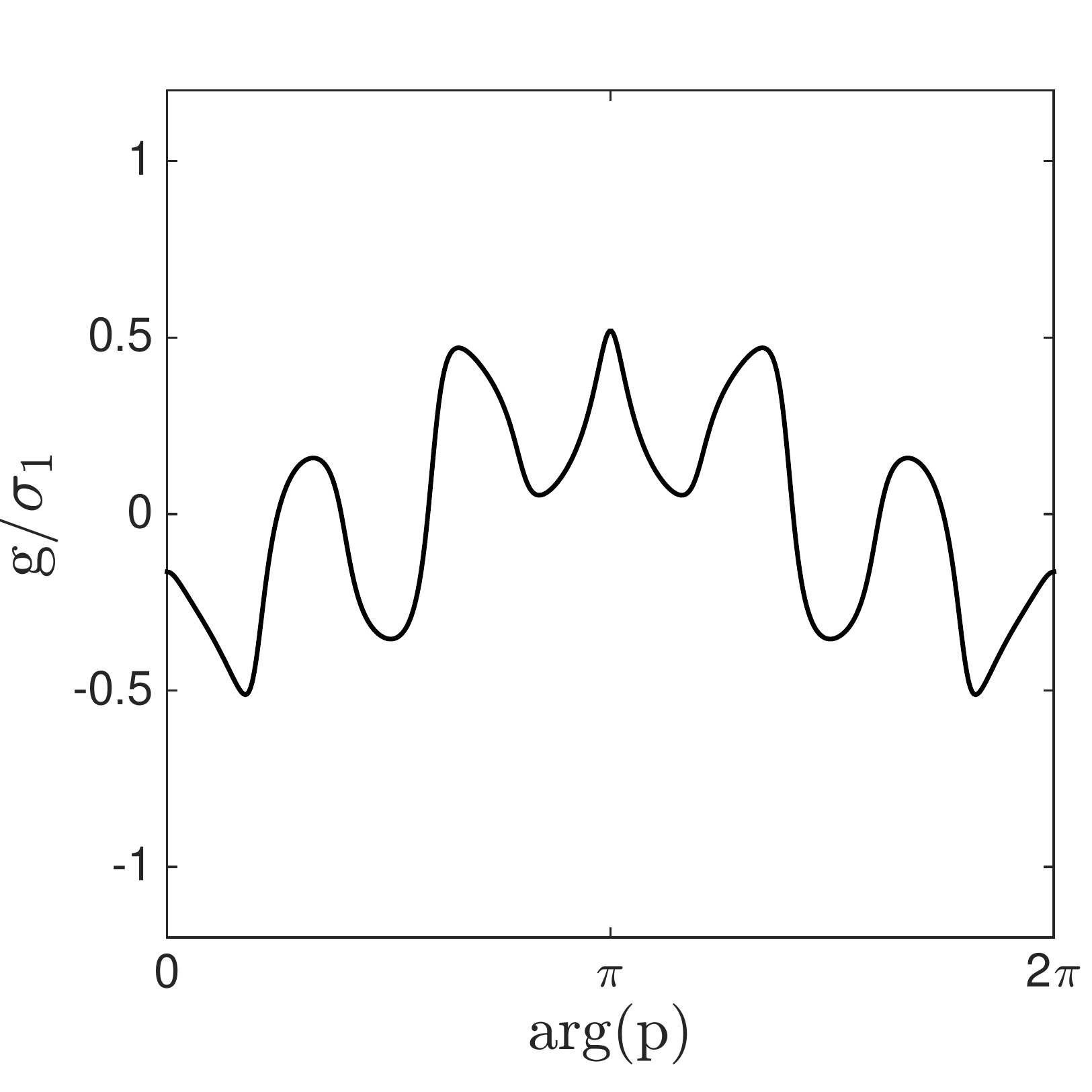}
      \caption{}
    \end{subfigure}
    \hskip .2cm
        \begin{subfigure}{0.4\textwidth}
      \hskip -2.5mm
      \includegraphics[height=4.4cm, width=5.1cm]{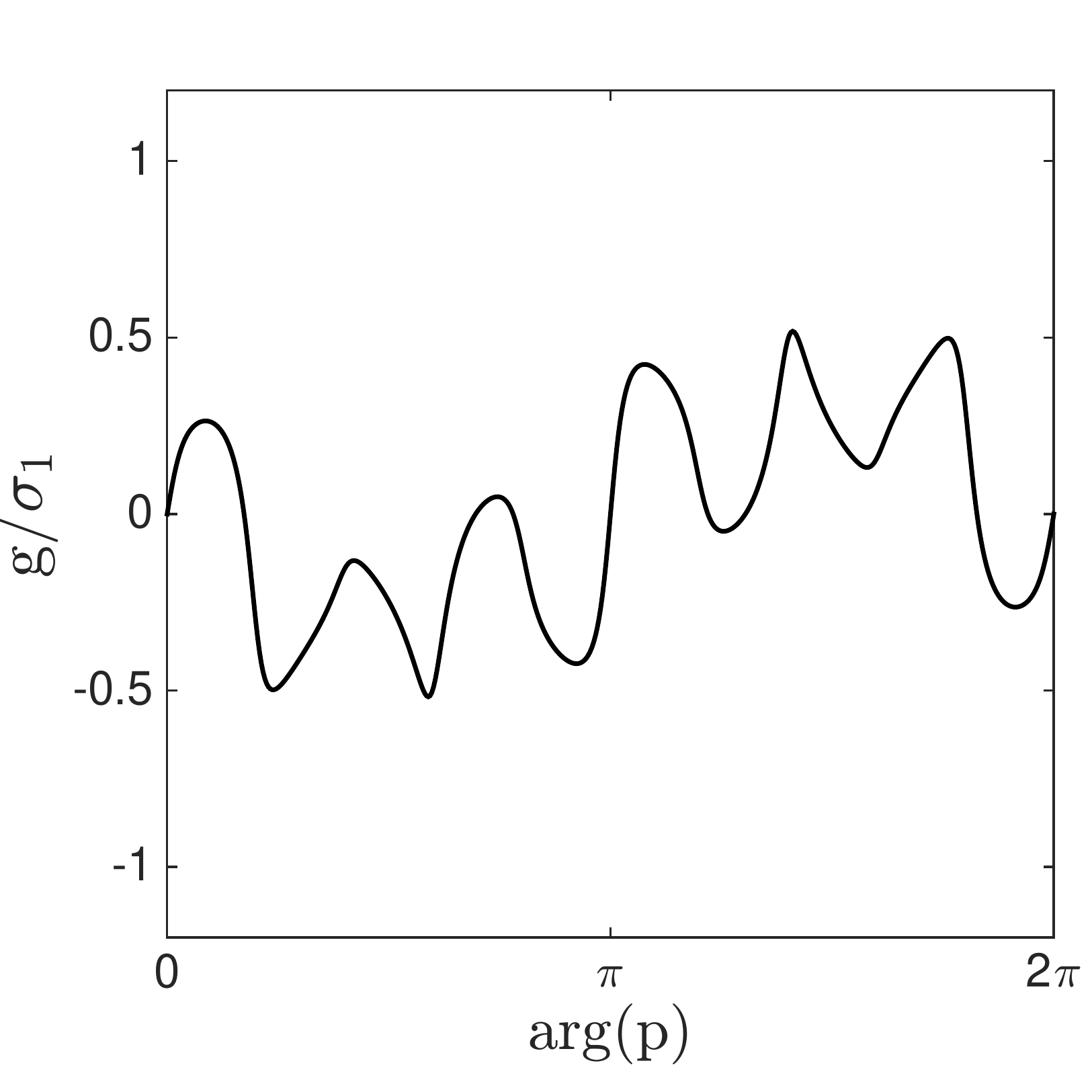}
      \caption{}
    \end{subfigure}
    \vskip -.3cm
     \begin{subfigure}{0.4\textwidth}
     \hskip -2.5mm
      \includegraphics[height=4.4cm, width=5.1cm]{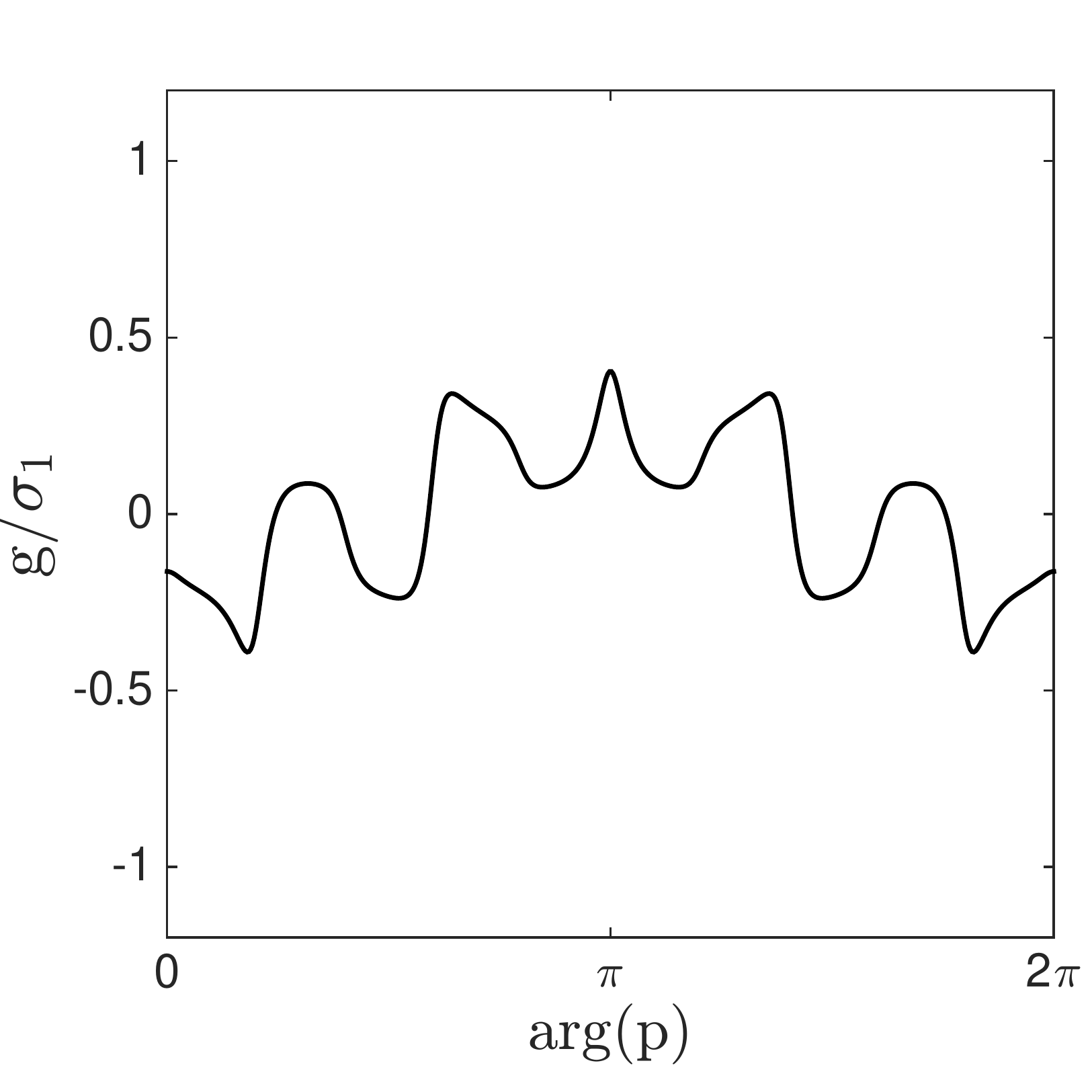}
      \caption{}
    \end{subfigure}
    \hskip .2cm
        \begin{subfigure}{0.4\textwidth}
      \hskip -2.5mm
      \includegraphics[height=4.4cm, width=5.1cm]{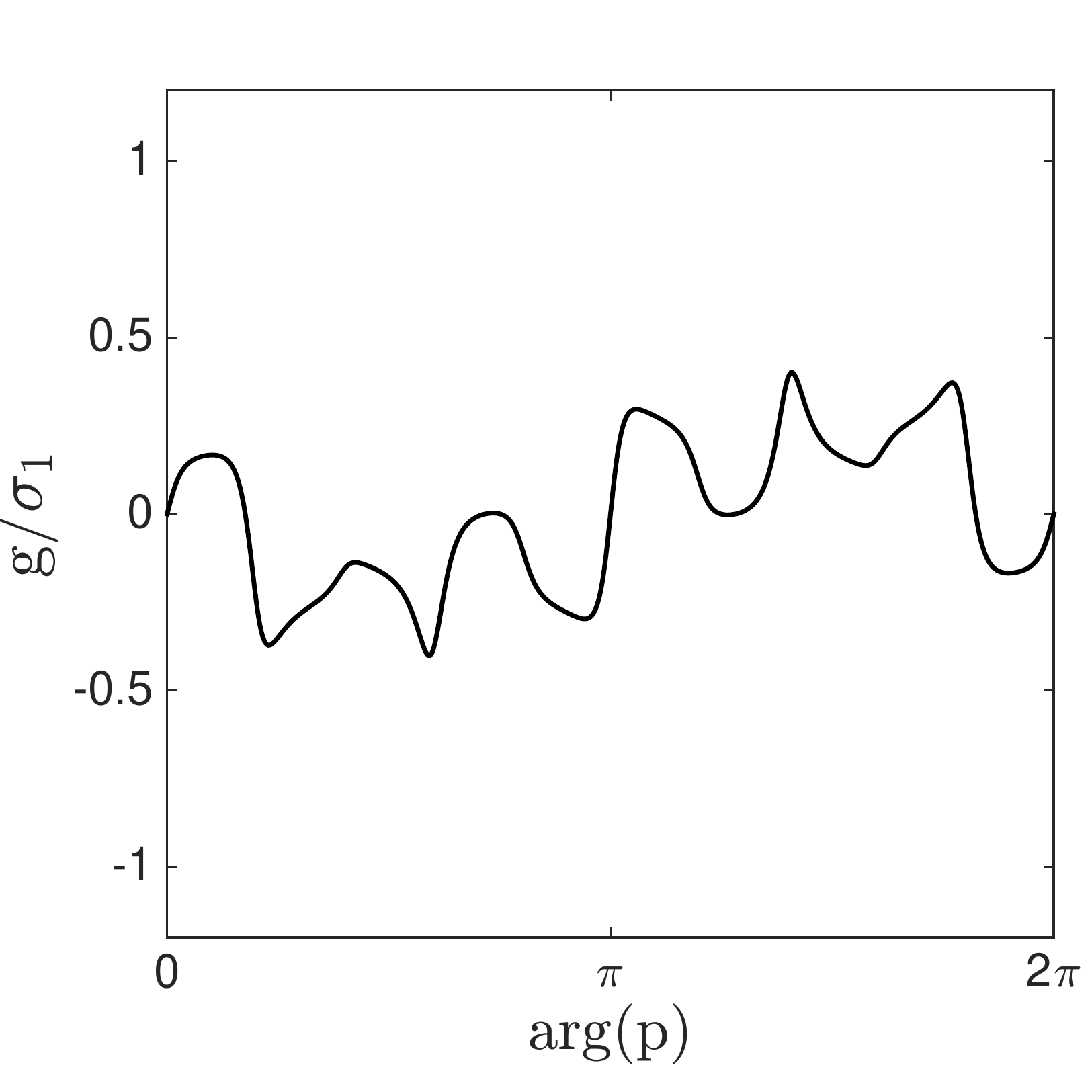}
      \caption{}
    \end{subfigure}
        \vskip -3mm
\caption{Flux function $g$ corresponding to the examples in Fig.\;\ref{fig:E_Om_star}.}\label{fig:E_Om_star_flux}
\end{figure}

Fig.\;\ref{fig:E_Om} and Fig.\;\ref{fig:E_Om_star} show $E_\Om$-inclusions obtained from the construction method described in section \ref{sec:E_Om;formulation}; the corresponding boundary flux $g$ on $\p \Om$ is drawn in Fig.\;\ref{fig:E_Om_flux} and Fig.\;\ref{fig:E_Om_star_flux}, respectively.  The pairs $(\Omega,D)$ and $g$ are given by \eqnref{Om:expression:2} and \eqnref{g:expression}.
These two examples clearly indicate that the same pair of regions $(\Omega,D)$ can induce an interior uniform field of multiple directions by choosing $g$ according to the direction of the uniform field.
 While most coefficients $a_n$ are zero for the examples in Fig.\;\ref{fig:E_Om}, the coefficients $a_n$ in Fig.\;\ref{fig:E_Om_star} decay relatively slowly as $n$ increases. The corresponding boundary flux $g$ in Fig.\;\ref{fig:E_Om_star_flux} is more oscillatory than that in Fig.\;\ref{fig:E_Om_flux}.

Examples in Fig.\;\ref{fig:E_Om_star} are created using the so-called 
Appell hypergeometric function
\begin{align*}
&F_1(a,b,b',c;x,y)\\
&=\sum_{m=0}^\infty\sum_{n=0}^\infty\frac{1}{m!n!}\frac{\Gamma(a+m+n)}{\Gamma(a)}\frac{\Gamma(b+m)}{\Gamma(b)}\frac{\Gamma(b'+n)}{\Gamma(b')}\frac{\Gamma(c)}{\Gamma(c+m+n)}x^m y^n.
\end{align*}
It is well known that $\Psi(z):=zF_1(1/5,4/5,-2/5,6/5,-z^5,z^5)$ maps the unit disc to a five-pointed star \cite{Brilleslyper:2012:ECA}. 
For the examples in Fig.\;\ref{fig:E_Om_star}, we set 
\beq\label{def:a_n}
a_n=
\begin{cases}
0.9^n*c_n,\quad&\mbox{for }1\leq n\leq 101,\\
0,\quad&\mbox{otherwise},
\end{cases}
\eeq 
where $c_n$ is the $z^n$-component coefficient of $\Psi(z)$. The coefficients $a_n$ exponentially decrease as $n$ increases (differently from $c_n$) so that the corresponding conformal mapping sends the unit disk to a smooth domain. Hence, $\Om$ has the shape of a polygon with rounded corners.

\section{Non-elliptical neutral coated inclusions in two dimensions}\label{sec:Neutral2D}
In this section, we present the construction of neutral coated inclusions in two dimensions by using an approach similar to that used in section \ref{sec:E_Omega} and \cite{Milton:2001:NCI}.
We now assume that the flux on the boundary of the core can be actively assigned. 
Previously \cite{Milton:2001:NCI}, neutral coated inclusions were constructed when the core was either a hole or a perfect conductor. We will see in section 4 that the analysis
  presented here is also relevant to the three-dimensional case where one seeks neutral coated inclusions having a geometry independent of $x_3$ and a coating that is anisotropic and with none of the
 crystal axes being aligned parallel to the $x_3$-axis. This provides additional motivation for studying it.

As in section \ref{sec:E_Omega}, $\Omega$ and $D$ are simply connected bounded planar domains such that $\overline{D}\subset\Om$. Let the coating phase $\Theta=\Omega\setminus\overline{D}$ have a constant isotropic conductivity $\sigma_1$. The exterior region $\RR^2\setminus\overline{\Om}$ is now occupied by a homogeneous material, possibly anisotropic, with the conductivity 
$${\bm{\sigma}}_0=\left( \begin{array}{cc}
\sigma_{11}&\sigma_{12} \\
\sigma_{12}& \sigma_{22}\\
 \end{array} \right).$$   
We consider the potential problem
\beq\label{eqn:trans1}
\begin{cases}
\ds\nabla\cdot{\bm{\sigma}}_0\nabla\varphi_0=0\quad& \mbox{in }\RR^2\setminus\overline{\Om} ,\\
\ds\Delta\varphi_1 = 0 \quad&\mbox{in }\Theta,\\
\ds\left(\bm{\sigma}_0\nabla\varphi_0\right)\cdot \Bn=\left(\sigma_1\nabla \varphi_1\right)\cdot \Bn\quad&\mbox{on }\p \Om,\\
\ds\varphi_0=\varphi_1\quad& \mbox{on }\p \Om
\end{cases}
\eeq  
with the flux condition on the boundary of the core
\beq\label{fluxcondition}
\ds \left(\sigma_1\nabla \varphi_1\right)\cdot \Bn=g\quad\mbox{on }\p D.
\eeq
Here, $g$ is a function whose integral over $\p D$ vanishes, and we assume it can be actively assigned depending on the exterior field.  
We further assume the uniformity condition
\beq\label{eqn:phi0_1}
\varphi_0(x_1,x_2)=-e_1 x_1 -e_2 x_2\quad\mbox{in }\RR^2\setminus\overline{\Om}\ \eeq
for some real constants $e_1$ and $e_2$.
We keep the notations \eqnref{ij_def} and \eqnref{ij_def2}. Note that, differently from section \ref{sec:E_Omega}, we set the uniformity condition exterior to $\Om$.

The problem \eqnref{eqn:trans1}-\eqnref{eqn:phi0_1} is over-determined, so that in general it has no solution.  For a given $\Om$, we construct the core $D$ and the flux function $g$ in the following subsections such that the problem \eqnref{eqn:trans1}-\eqnref{eqn:phi0_1} admits a solution. 
For such a case, the coated inclusion $\Om$ does not perturb the exterior uniform field $\varphi_0$. In other words, it is neutral to $\varphi_0$.

\subsection{Analytic function formulation}\label{sec:complexpotential}
As in section \ref{sec:E_Omega},  we reformulate the over-determined problem \eqnref{eqn:trans1}-\eqnref{eqn:phi0_1} by following the complex potential approach in \cite{Milton:2001:NCI}.

As $\varphi_1$ is harmonic in the doubly connected domain $\Theta$ and has a mean-zero normal flux on $\p D$, it admits a complex analytic function $$w(z)=\varphi_1(z)+\iu\psi_1(z)\quad\mbox{in }\Theta.$$
From the Cauchy-Riemann equations, we have
\beq\label{Psi_tau_2}
\pd{\psi_1}{\Bt}=\pd{\varphi_1}{\Bn}\quad\mbox{on }\p \Om,\p D.
\eeq
We then obtain from \eqnref{fluxcondition} that
\beq  \label{eqn:psi1_1}
\pd{\psi_1}{\Bt} =\frac{1}{\sigma_1}g\quad\mbox{on }\p D.
\eeq
The relations \eqnref{eqn:trans1}, \eqnref{eqn:phi0_1} and \eqnref{Psi_tau_2} imply
\beq
\pd{\psi_1}{\Bt}=\pd{\varphi_1}{\Bn} =\frac{1}{\sigma_1}\left(\bm{\sigma}_0\nabla\varphi_0\right)\cdot \Bn
=\frac{1}{\sigma_1}(-\mathbf{j}_0)\cdot \Bn = \frac{1}{\sigma_1}(j_2,-j_1)\cdot\Bt\quad\mbox{on }\p\Om.
\eeq
Hence, we have (the constant term is neglected)
\beq \label{eqn:psi1_2}
\psi_1(z)=\frac{1}{\sigma_1}\left(j_2x_1-j_1x_2\right)\quad\mbox{on }\p\Om.
\eeq
Using this relation together with \eqnref{eqn:phi0_1}, one can easily derive the relation
\beq\label{wz_1}
w(z)=kz+h\bar{z}\quad\mbox{on }\p\Om
\eeq
with the complex constants $k$ and $h$ given by \eqnref{gh}.

As discussed in section \ref{sec:E_Omega}, there is a conformal mapping, namely $z(p)$, from an annulus $\{p:r<|p|<R\}$ to $\Theta$ for some $0<r<R$ and the functions $z(p)$ and $w(p):=w(z(p))$ admit the Laurent series expansions
\begin{align}
z(p)&=\sum_{n=-\infty}^\infty a_n p^n,\\
w(p)&=\sum_{n=-\infty}^{\infty}b_n p^n
\end{align}
for $r<|p|<R$ with some complex coefficients $a_n$ and $b_n$. The coefficients $a_n$ are associated with $\Theta$, and the coefficients $b_n$ should be determined by $a_n$ and $\varphi_0$ such that the boundary relation \eqnref{wz_1} holds. In other words,
\beq\label{eqn:bn_1}
b_n = k a_n + hR^{-2n}\overline{a_{-n}}\quad\mbox{for all } n\in\ZZ.
\eeq

We can construct active neutral inclusions by specifying the coefficients $a_n$ as follows. 
We first choose the geometric coefficients $a_n$ and set the pair of domains $(\Om, D)$ such that 
\beq\label{Om:expression:2:neutral}
\p \Om = \{z(p):|p|=R\},\quad \p D =\{z(p):|p|=r\}.
\eeq
We then determine $b_n$ by \eqnref{eqn:bn_1}, for a given arbitrary uniform field $\varphi_0$. Given that $w(p)$ converges to an analytic function in $\{p:r<|p|<R\}$, the function 
\beq\label{varphi1:w:2}
\varphi_1(z)=\Re\{w(z)\},\quad z\in\Om\setminus\overline{D},
\eeq
satisfies the over-determined problem \eqnref{eqn:trans1}-\eqnref{eqn:phi0_1} with
\beq\label{g:expression:2}
g =\sigma_1 \pd{\Im\{w\}}{\Bt}\quad\mbox{on }\p D. 
\eeq
As discussed in section \ref{sec:E_Omega}, the convergence of $w(p)$ is independent of the direction of the uniform field $\varphi_0$. Therefore, the constructed pair of domains $(\Om, D)$ is neutral to the arbitrary uniform field $\varphi_0$, where the flux on $\p D$ is actively assigned depending on $\varphi_0$.

\captionsetup[subfigure]{labelformat=empty}
 \begin{figure}[p]
 \centering
   \begin{subfigure}{0.4\textwidth}
      \centering
      \includegraphics[height=4cm]{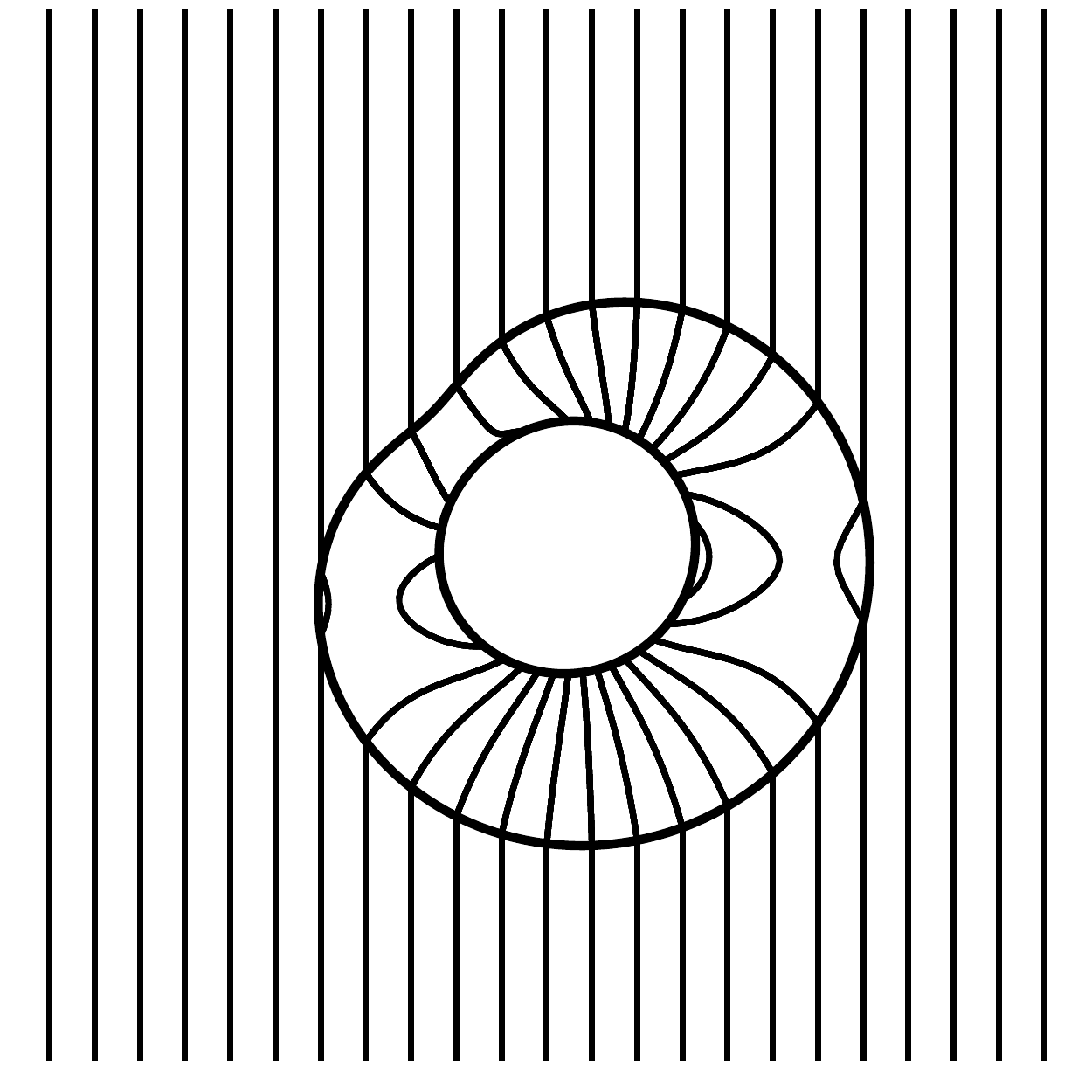}
      \caption{(a)}
    \end{subfigure}
       \begin{subfigure}{0.4\textwidth}
      \centering
      \includegraphics[height=4cm]{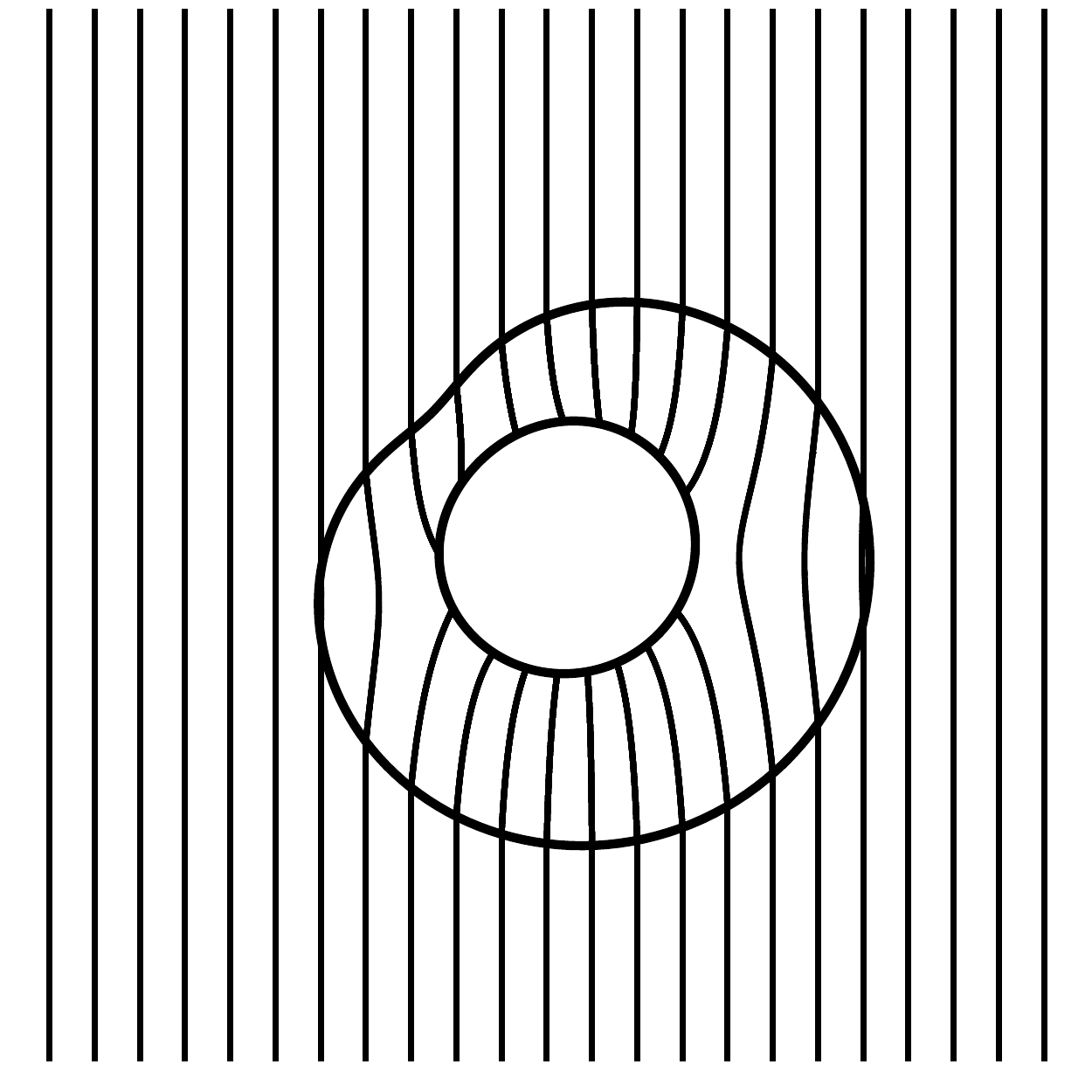}
      \caption{(b)}
    \end{subfigure}
    \begin{subfigure}{0.4\textwidth}
      \centering
      \includegraphics[height=4cm]{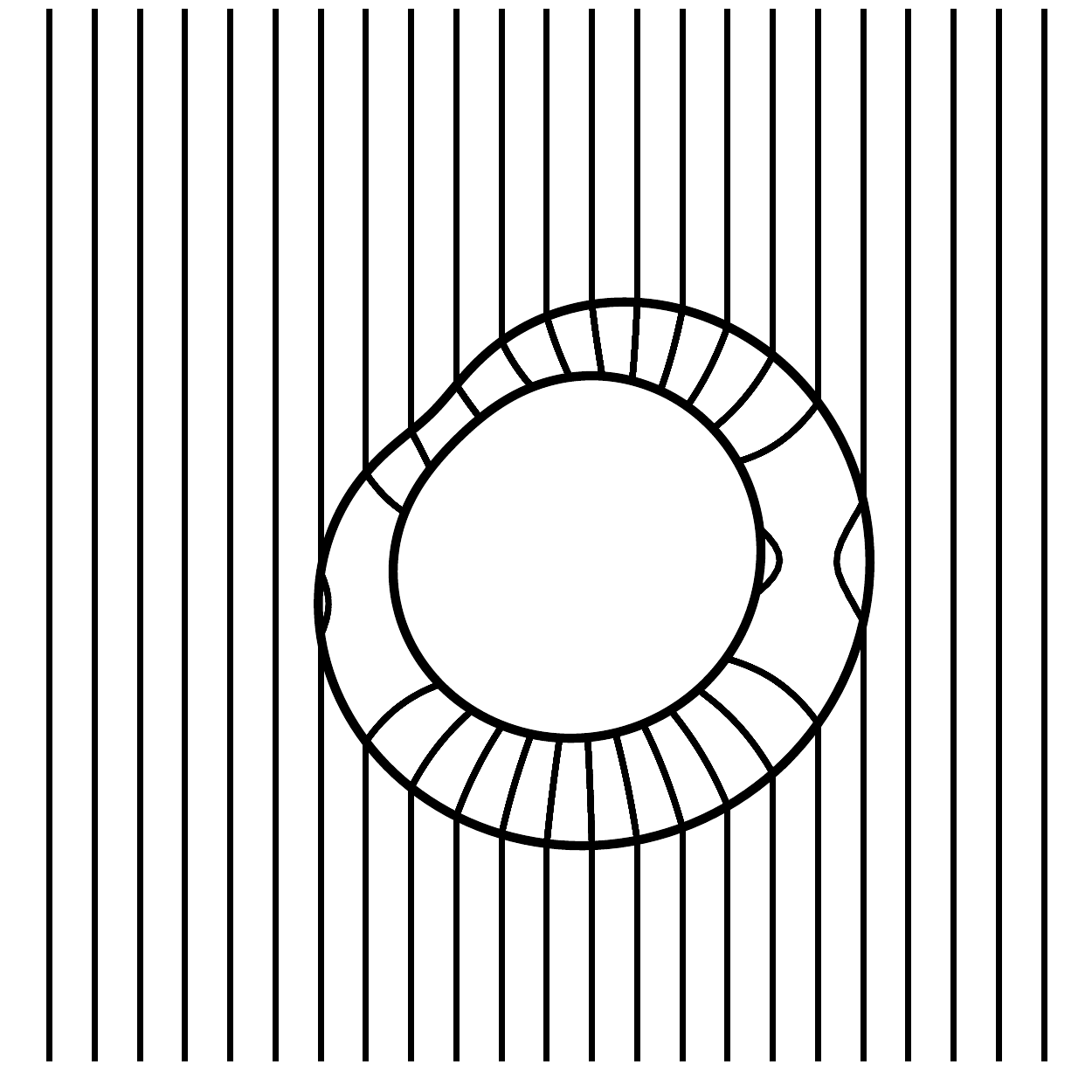}
      \caption{(c)}
    \end{subfigure}
        \begin{subfigure}{0.4\textwidth}
      \centering
      \includegraphics[height=4cm]{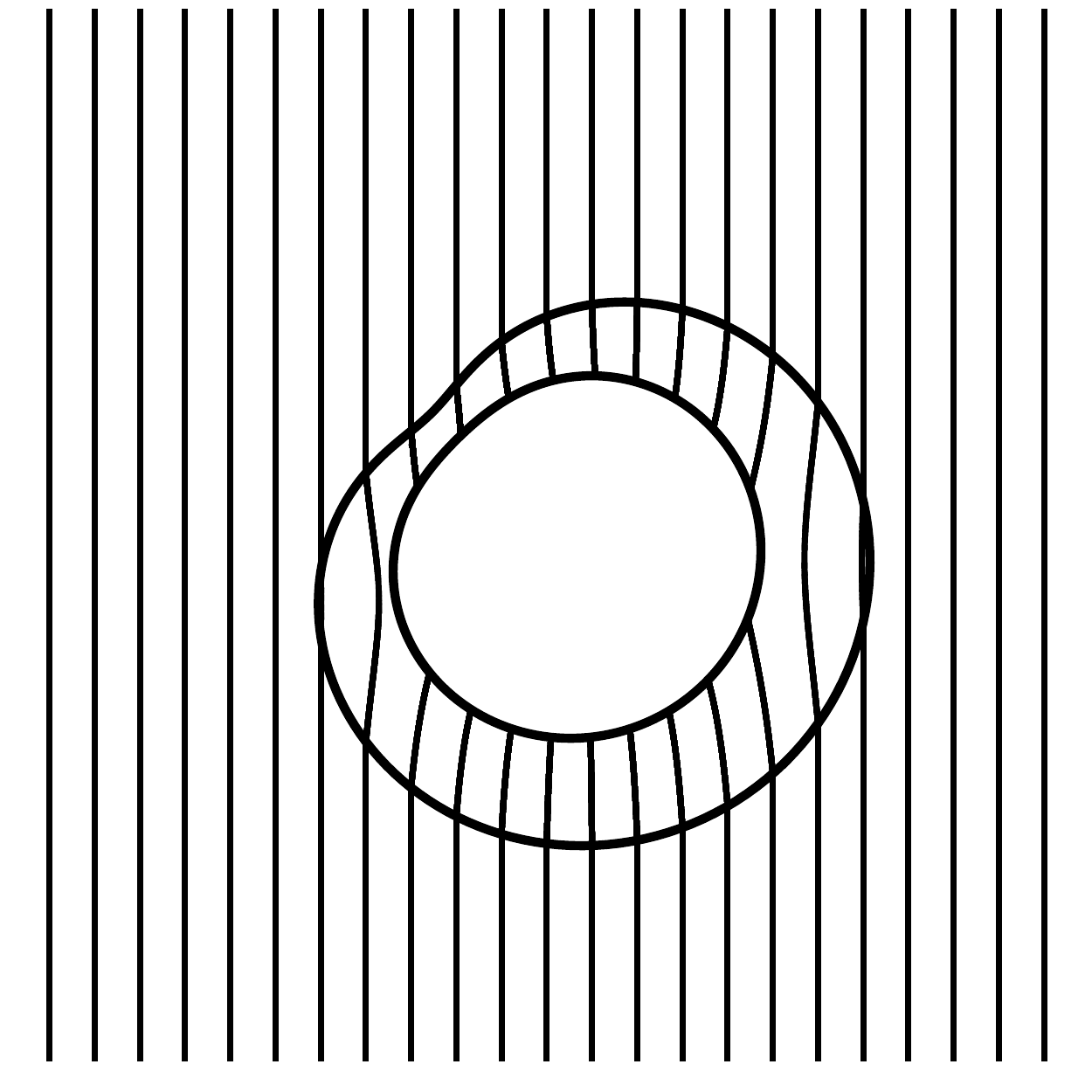}
      \caption{(d)}
    \end{subfigure}
   \vskip -1mm
\caption{Non-elliptical active neutral inclusions. The figures illustrate $\p D$, $\p\Om$, and the current flow in $\RR^2\setminus \overline{D}$ (equipotential lines of $\varphi_1$).
The non-zero geometric parameters are $a_1 = 1-\iu,\ a_2=0.3-0.3\iu$. We set $R=1$. The conductivity $\sigma_1$ of the coating and the parameter $r$ are as follows: (a) $\sigma_1/\sigma_0=4,\ r=0.5$; (b) $\sigma_1/\sigma_0=1.22,\ r=0.5$; (c) $\sigma_1/\sigma_0=4,\ r=0.7$;  (d) $\sigma_1/\sigma_0=1.22,\ r=0.7$.}\label{fig:active_cloaking}

 \centering
   \begin{subfigure}{0.4\textwidth}
      \hskip -2.5mm
      \includegraphics[height=4.4cm, width=5.1cm]{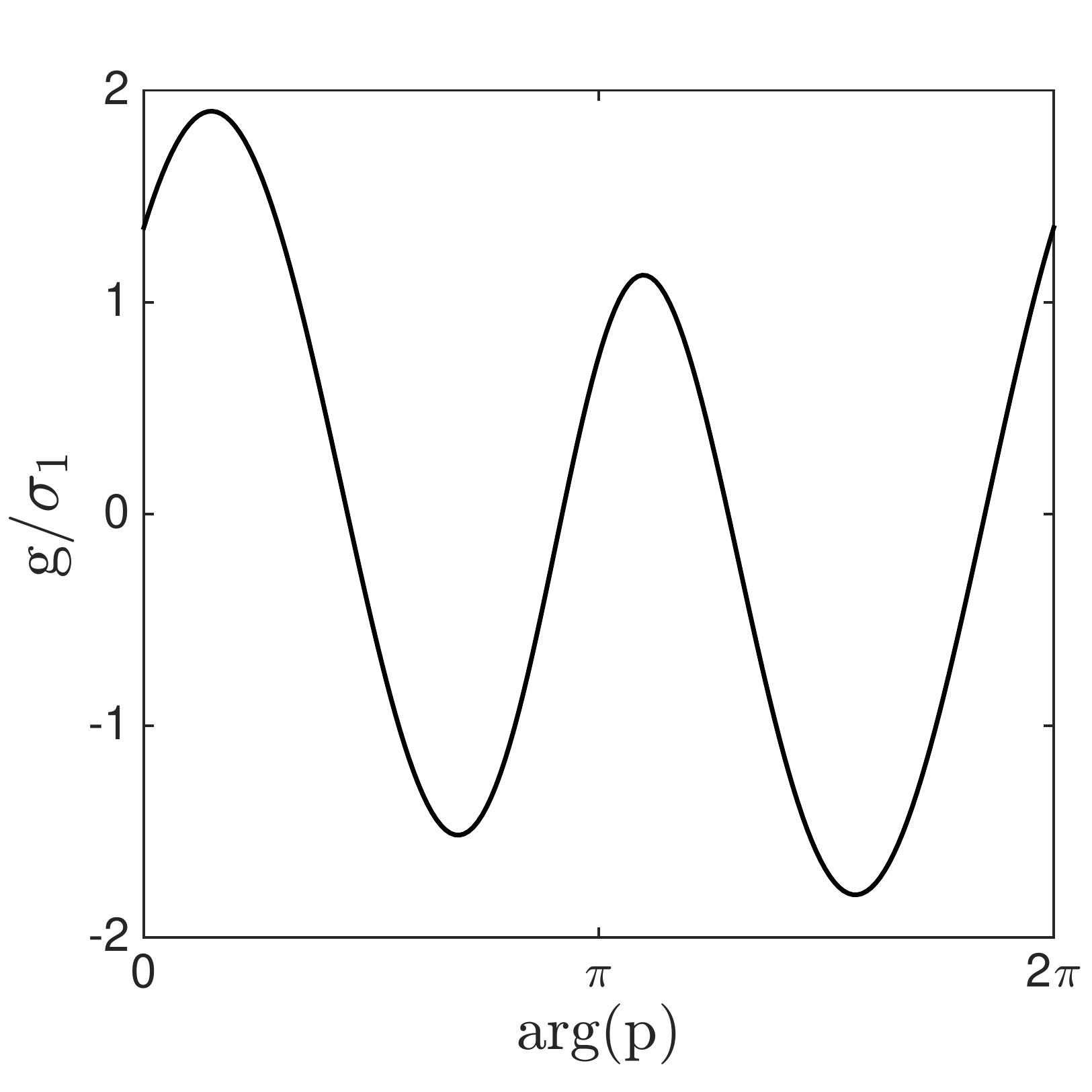}
      \caption{(a)}
    \end{subfigure}
    \hskip .2cm
       \begin{subfigure}{0.4\textwidth}
      \hskip -2.5mm
      \includegraphics[height=4.4cm, width=5.1cm]{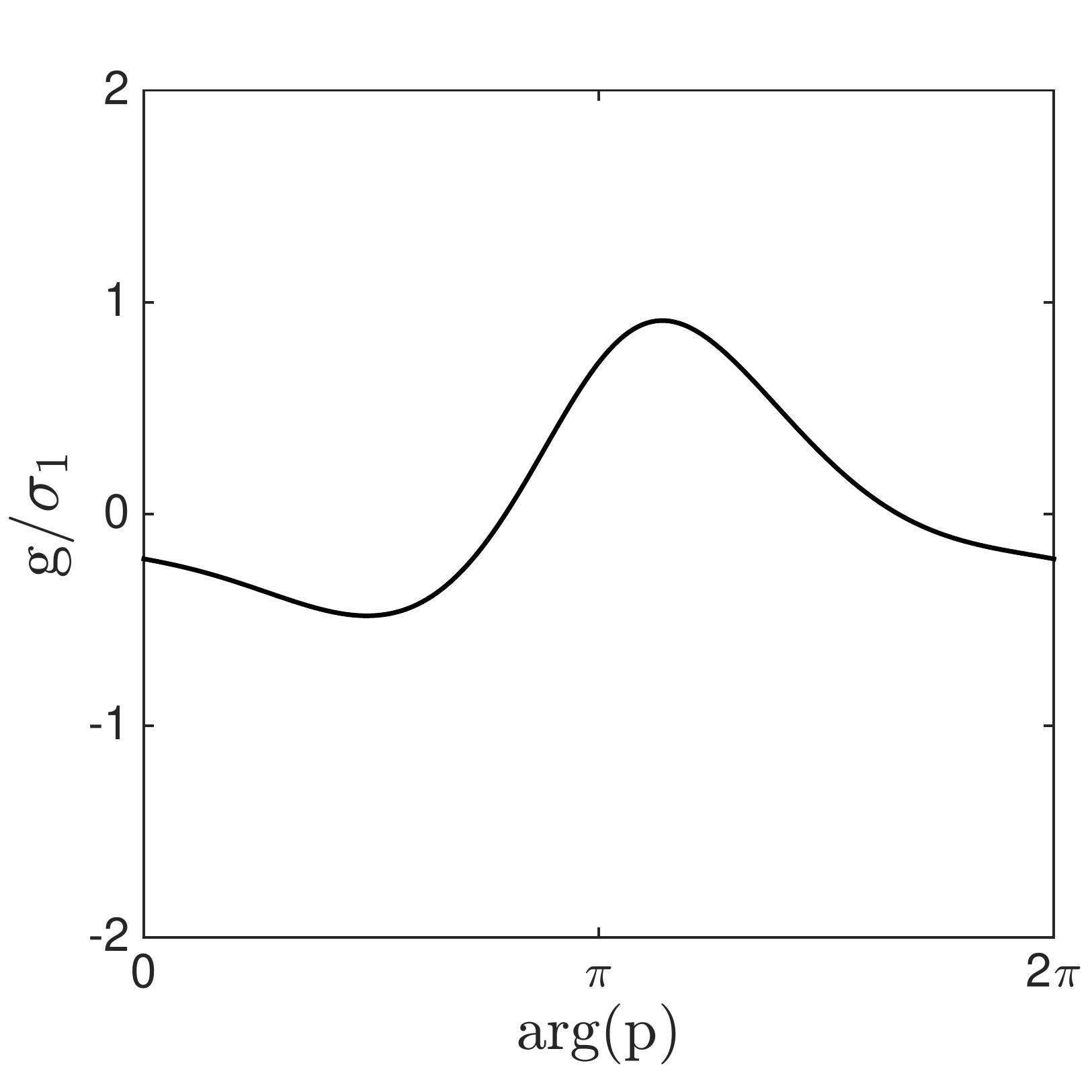}
      \caption{(b)}
    \end{subfigure}
    \vskip -.3cm
    \begin{subfigure}{0.4\textwidth}
     \hskip -2.5mm
      \includegraphics[height=4.4cm, width=5.1cm]{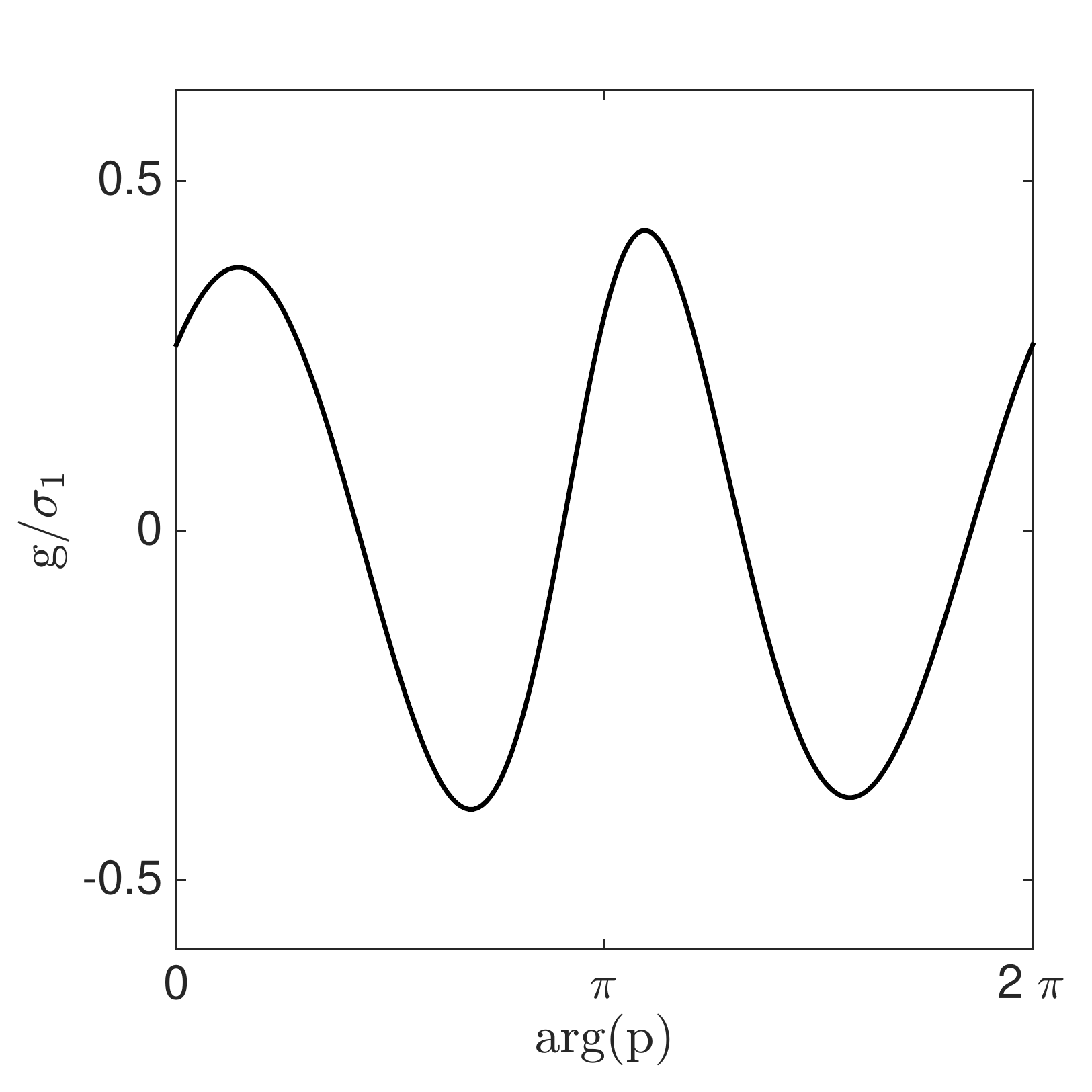}
      \caption{(c)}
    \end{subfigure}
    \hskip .2cm
        \begin{subfigure}{0.4\textwidth}
     \hskip -2.5mm
      \includegraphics[height=4.4cm, width=5.1cm]{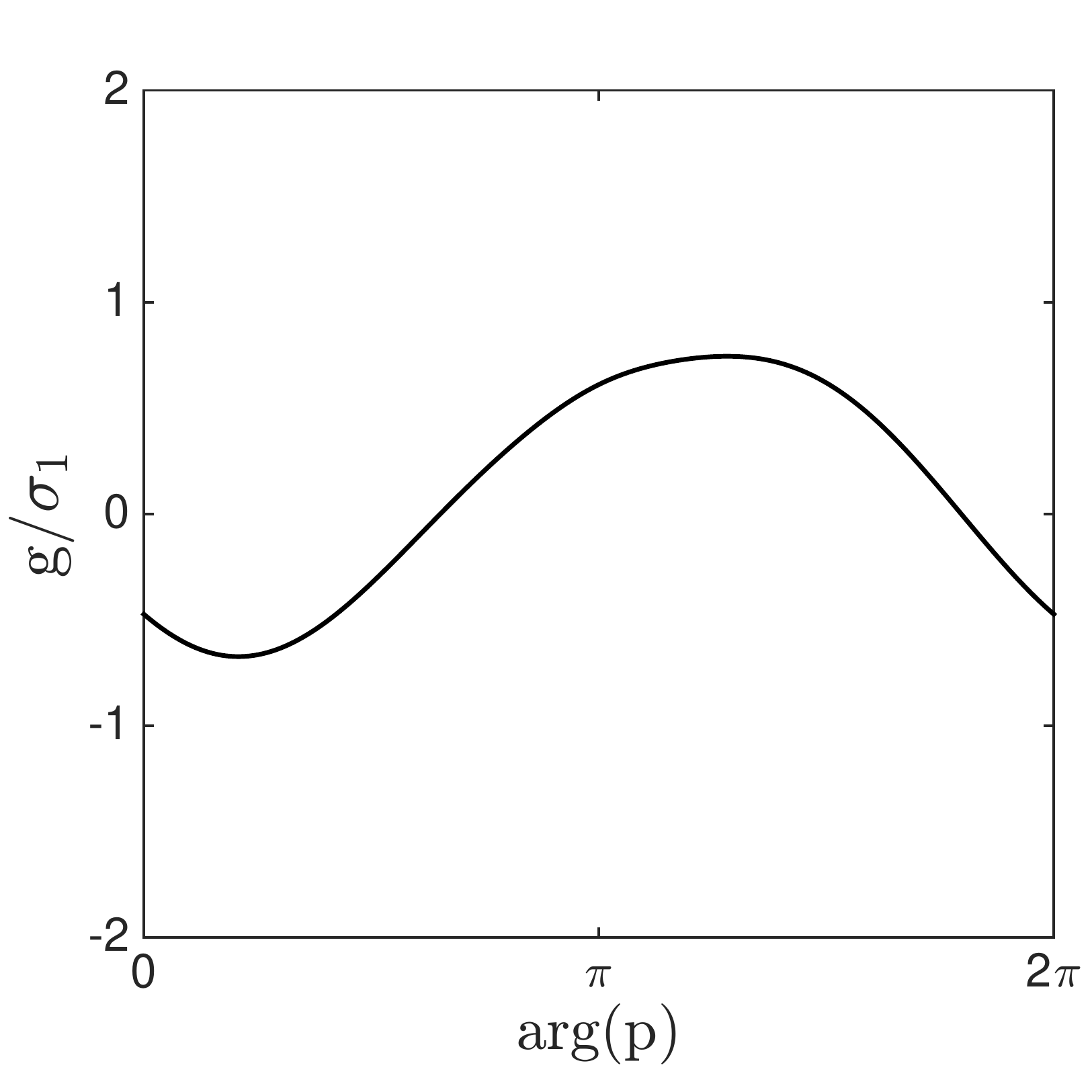}
      \caption{(c)}
    \end{subfigure}
        \vskip -3mm
\caption{Boundary flux $g$ on $\p D$ that corresponds to the examples in Fig.\;\ref{fig:active_cloaking}.}\label{fig:active_cloaking_flux}
\end{figure}

\captionsetup[subfigure]{labelformat=empty}
 \begin{figure}[p]
 \centering
   \begin{subfigure}{0.4\textwidth}
         \centering
      \includegraphics[height=4cm]{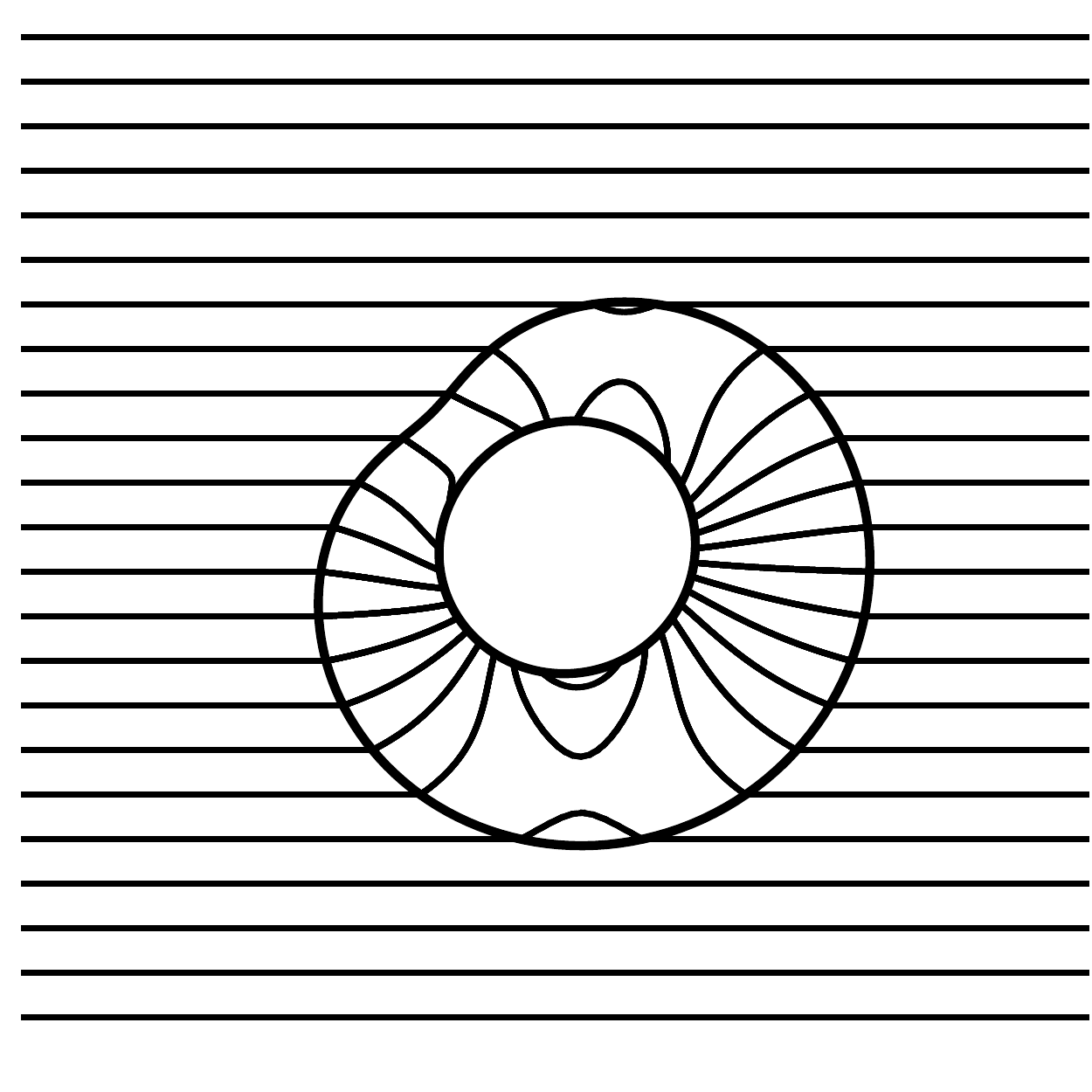}
      \caption{(a) }
    \end{subfigure}
       \begin{subfigure}{0.4\textwidth}
      \centering
      \includegraphics[height=4cm]{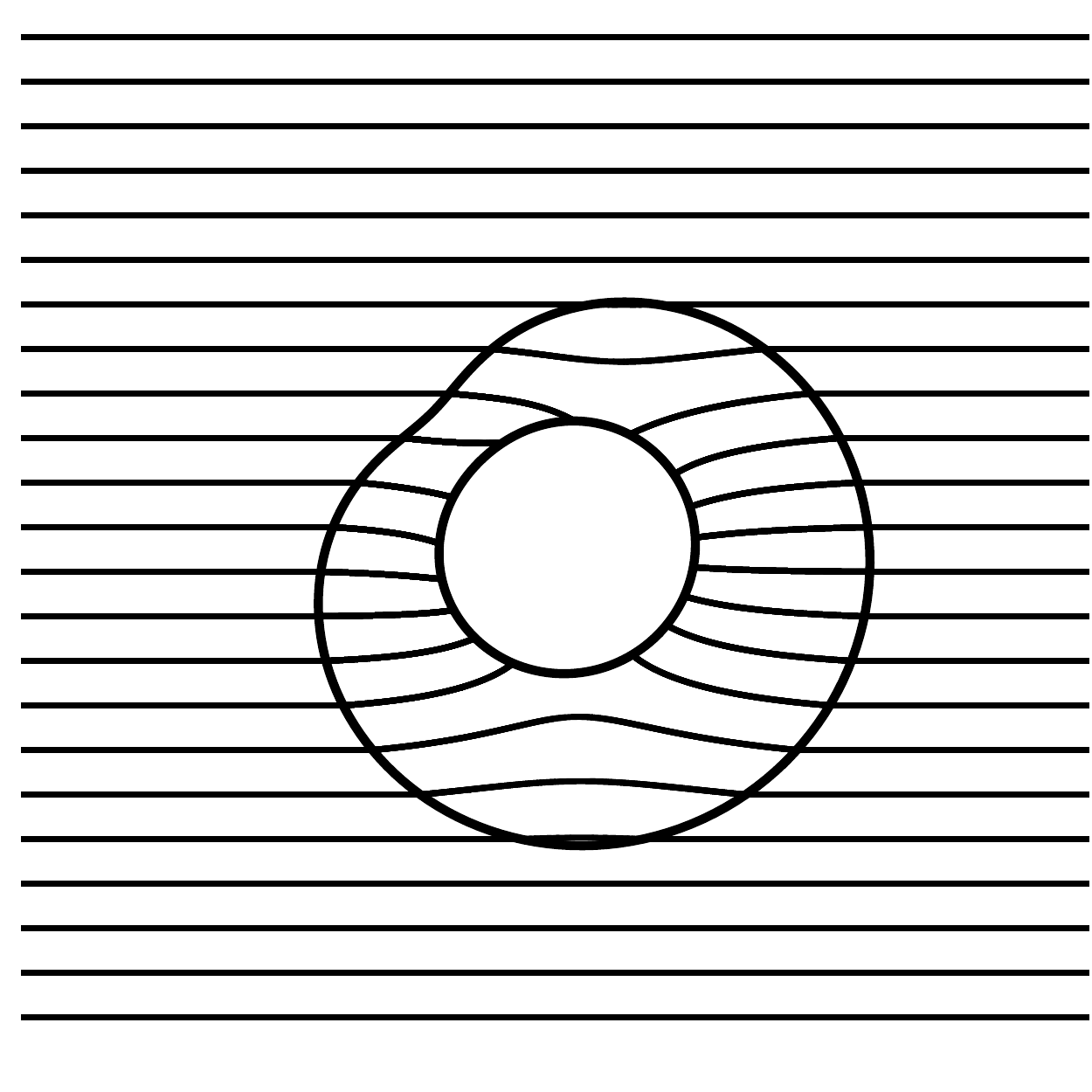}
      \caption{(b)}
    \end{subfigure}
    \begin{subfigure}{0.4\textwidth}
      \centering
      \includegraphics[height=4cm]{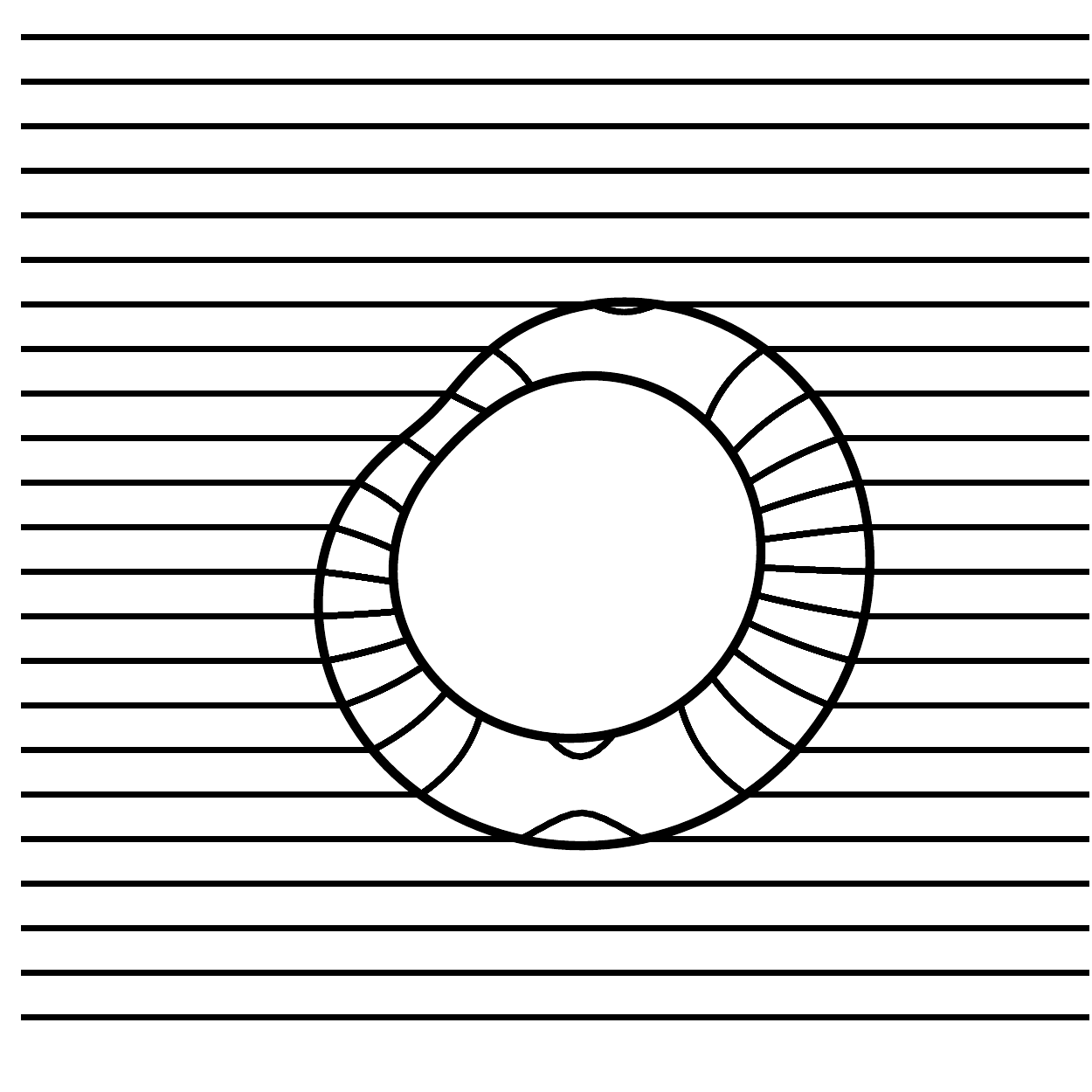}
      \caption{(c)}
    \end{subfigure}
        \begin{subfigure}{0.4\textwidth}
      \centering
      \includegraphics[height=4cm]{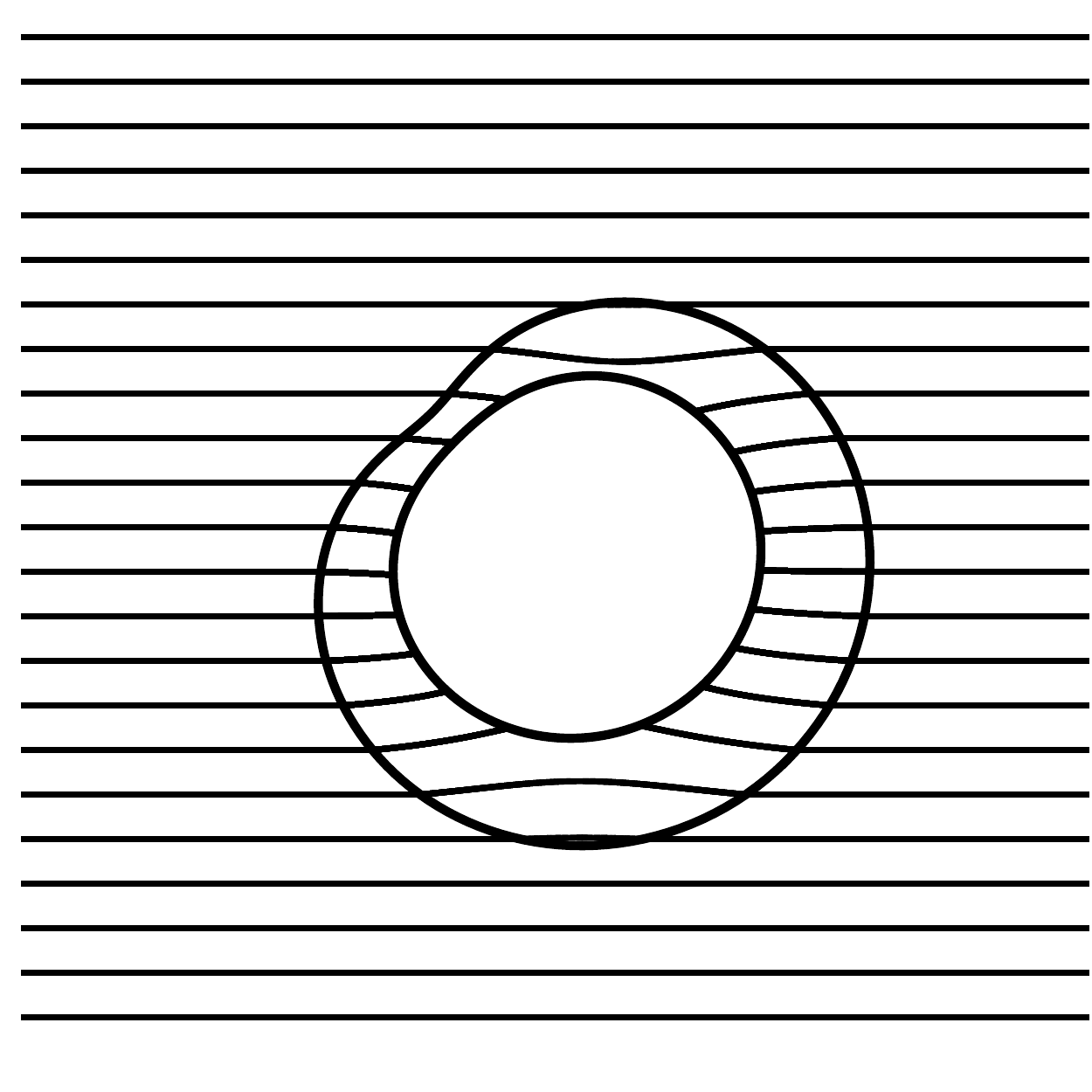}
      \caption{(d)}
    \end{subfigure}
      \vskip -1mm
\caption{Non-elliptical active neutral inclusions. The figures illustrate $\p D$, $\p\Om$, and the current flow in $\RR^2\setminus \overline{D}$ (equipotential lines of $\varphi_1$). The geometry and material parameters ($a_n$, $r$, and $\sigma_1/\sigma_0$) are the same as in Fig.\,\ref{fig:active_cloaking}; however, the direction of the current flow differs from that in Fig.\;\ref{fig:active_cloaking}.
}\label{fig:active_cloaking_2}

 \centering
  \begin{subfigure}{0.4\textwidth}
      \hskip -2.5mm
      \includegraphics[height=4.4cm, width=5.1cm]{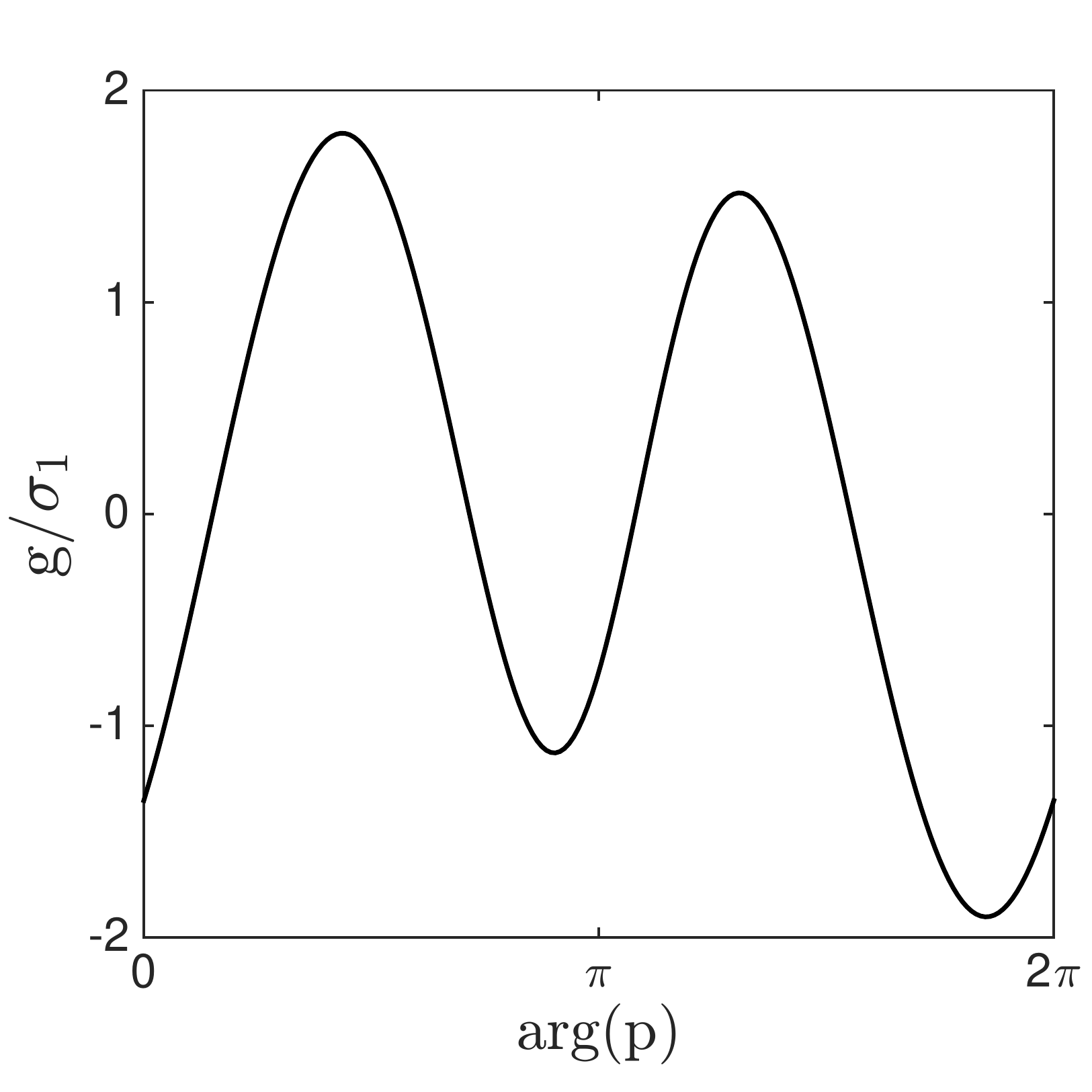}
      \caption{(a)}
    \end{subfigure}
    \hskip .2cm
       \begin{subfigure}{0.4\textwidth}
      \hskip -2.5mm
      \includegraphics[height=4.4cm, width=5.1cm]{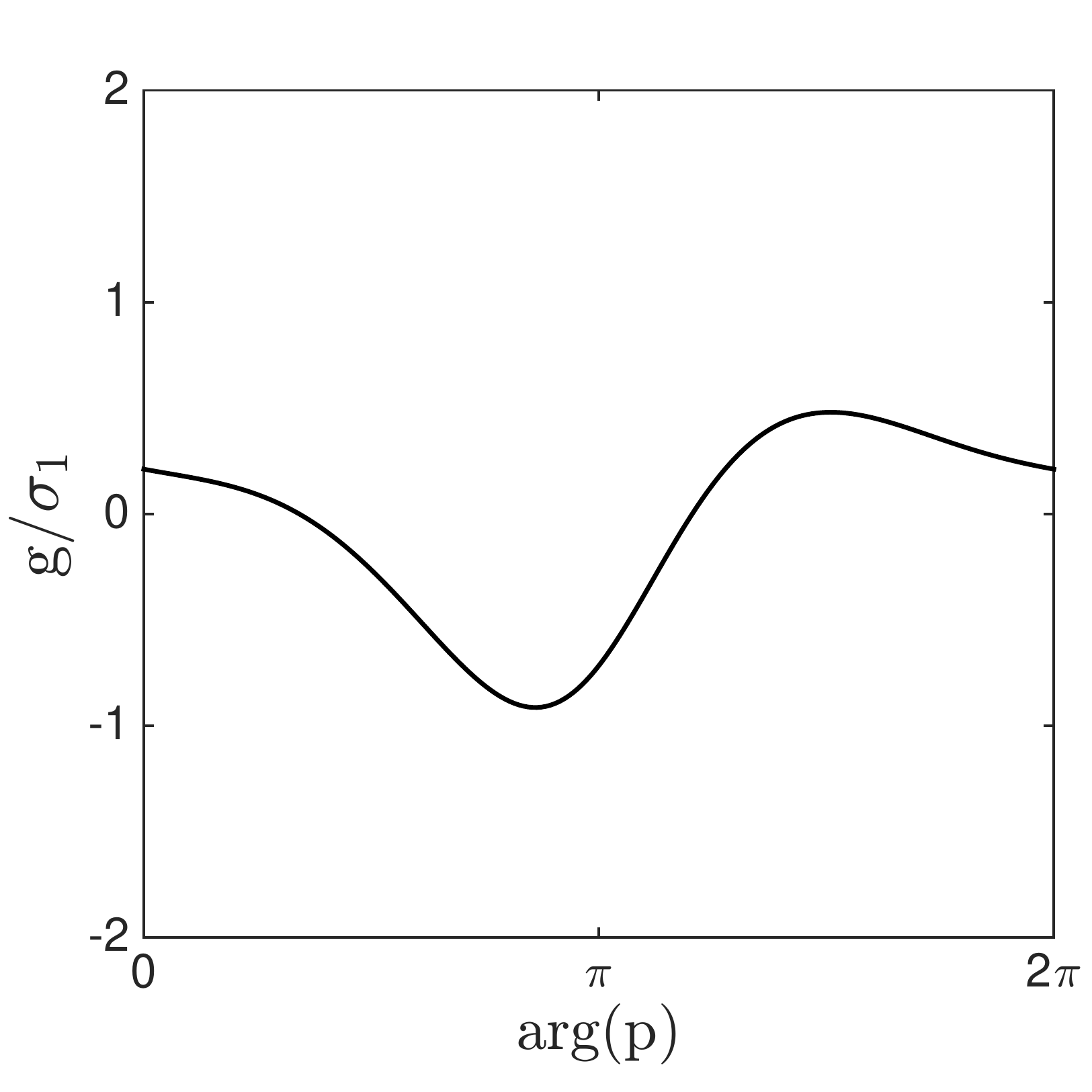}
      \caption{(b)}
    \end{subfigure}
    \begin{subfigure}{0.4\textwidth}
      \hskip -2.5mm
      \includegraphics[height=4.4cm, width=5.1cm]{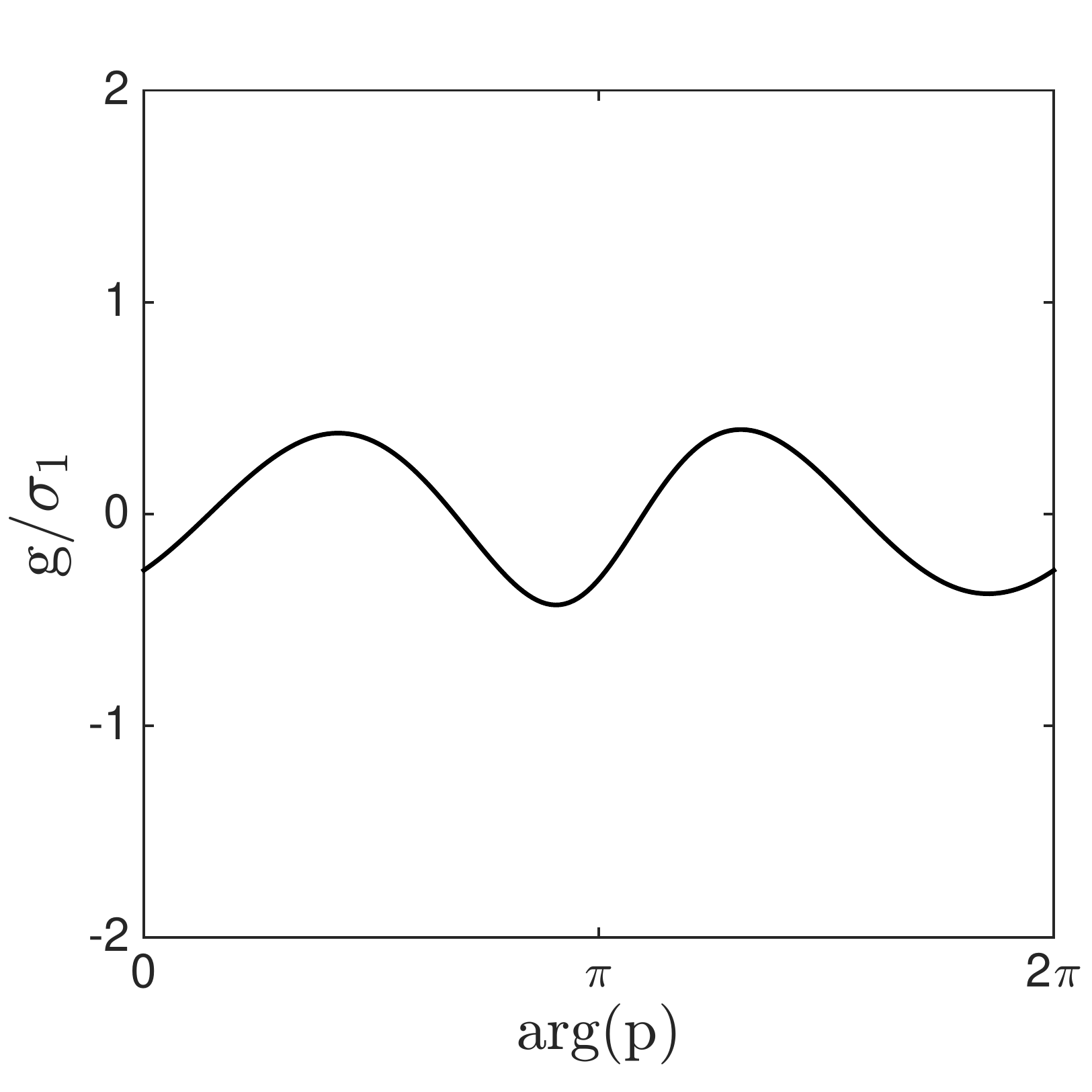}
      \caption{(c)}
    \end{subfigure}
    \hskip .2cm
        \begin{subfigure}{0.4\textwidth}
      \hskip -2.5mm
      \includegraphics[height=4.4cm, width=5.1cm]{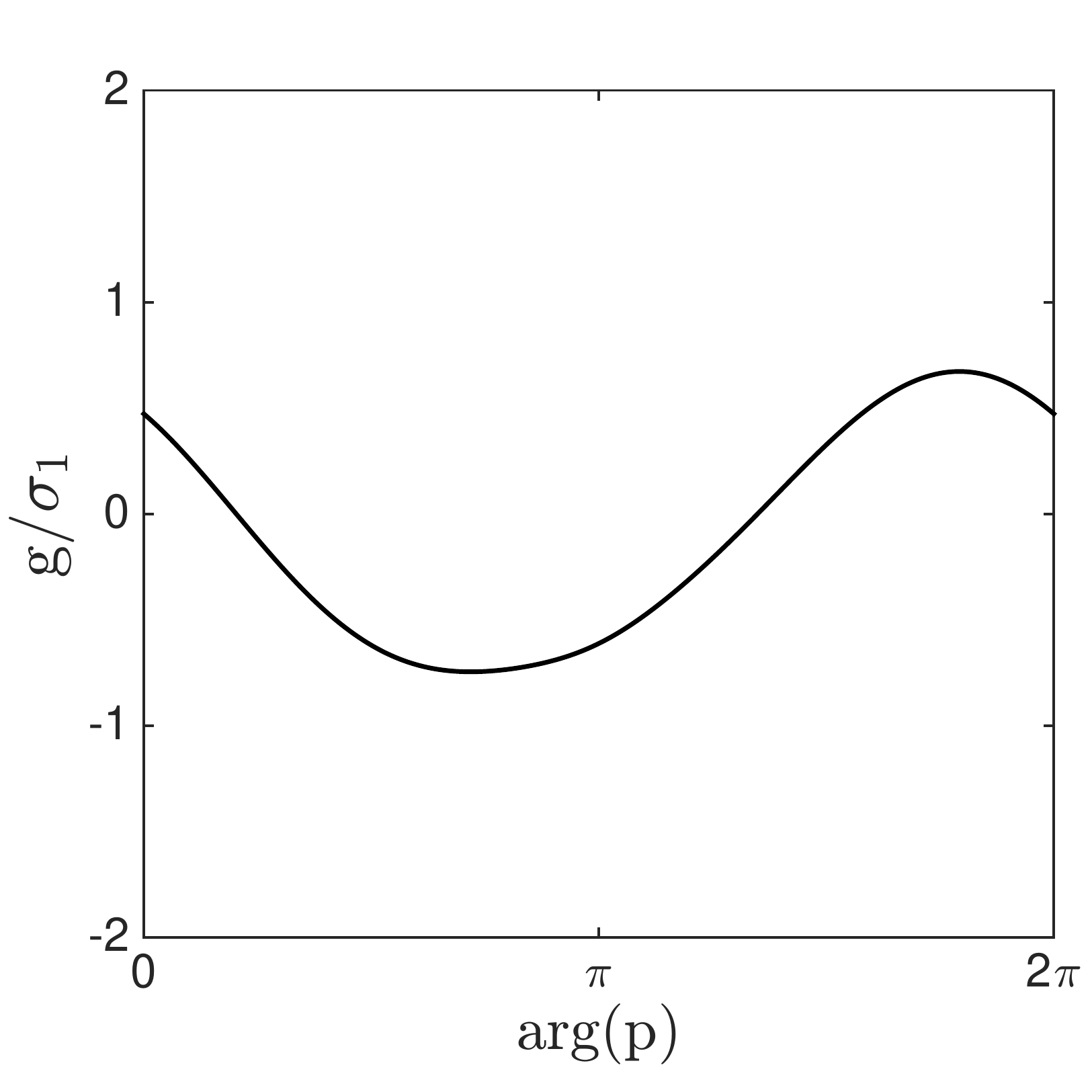}
      \caption{(d)}
\end{subfigure}
     \vskip -3mm
\caption{Boundary flux $g$ on $\p D$ that corresponds to the examples in Fig.\;\ref{fig:active_cloaking_2}.}\label{fig:active_cloaking_flux_2}\end{figure}

\subsection{Neutral coated inclusion with an outer domain $\Om$ of arbitrary shape}\label{sec:Gauert}
The proposed construction scheme makes it possible to find a neutral coated inclusion  with an outer domain $\Om$ of arbitrary shape.
Let $\Om$ be an arbitrary simply connected domain and denote $$z(p)=\sum_{n=0}^{+\infty} a_n p^n$$ a conformal mapping from the unit disk to $\Om$. 
{We give an additional regularity assumption on $D$ that $z(p)$ is analytic and univalent in $\{p:|p|<\rho^*\}$ for some $\rho^*>1$.}
Let us fix an arbitrary number $r$ satisfying $1/\rho^*<r<1$ and $R=1$. We set $D$ by \eqnref{Om:expression:2:neutral}. Then, for a given arbitrary uniform potential $\varphi_0$, the resulting Laurent series 
$$w(p)=\sum_{n=0}^{+\infty} ka_n p^n +\sum_{n=-\infty}^0 h\overline{a_{-n}}p^n$$
is analytic in $\{p\in\CC:1/\rho^*<|p|<\rho^*\}$.
Then, $\varphi_1(z):=\Re\{w(z)\}$ satisfies the over-determined problem \eqnref{eqn:trans1}-\eqnref{eqn:phi0_1} with $g$ given by \eqnref{g:expression:2}. 
In other words, the coated inclusion $(\Om,D)$ with the flux condition $g$ on $\p D$ is neutral to $\varphi_0$.

We would like to emphasize that the values of $k$ and $h$ can be assigned such that they are appropriate for a given $\varphi_0$. In other words, the pair of domains $(\Omega,D)$ is neutral to any external uniform field, where the flux condition on the boundary of the core is suitably chosen depending on the direction of the uniform field.

\subsection{Numerical Examples} 

Fig.\;\ref{fig:active_cloaking} and Fig.\;\ref{fig:active_cloaking_2} illustrate active neutral inclusions. The pair $(\Omega,D)$ and the boundary flux $g$ are constructed based on the conformal mapping expression \eqnref{Om:expression:2:neutral} and \eqnref{g:expression:2}. The corresponding boundary flux $g$ on $\p D$ is shown in Fig.\;\ref{fig:active_cloaking_flux} and in Fig.\;\ref{fig:active_cloaking_flux_2}, respectively. 
Although the pairs $(\Omega,D)$ are exactly the same in the two figures Fig.\;\ref{fig:active_cloaking} and Fig.\;\ref{fig:active_cloaking_2}, the background potential generates a horizontal current flow in Fig.\;\ref{fig:active_cloaking} and a vertical current flow in Fig.\;\ref{fig:active_cloaking_2}; furthermore, the flux $g$ is defined in accordance with the exterior field.

\section{Neutral cylindrical inclusions in three dimensions}\label{sec:3D}
We now consider a cylindrical region $(\Om\setminus \overline{D})\times \RR$, where 
 $\Omega$ and $D$ are simply connected planar domains satisfying $\overline{D}\subset \Om$. 
We set $\Theta=\Om\setminus\overline{D}$ as in the previous sections and denote the conductivity in the coating phase $\Theta\times \RR$ by 
 $$\bm{\sigma}=\bm{\sigma}(x_1,x_2,x_3)=
\left( \begin{array}{ccc}
\sigma_{11} & \sigma_{12} & \sigma_{13}\\
\sigma_{12} & \sigma_{22} & \sigma_{23}\\
 \sigma_{13} & \sigma_{23} & \sigma_{33}\end{array} \right).
$$
The matrix $\bm{\sigma}$ is assumed to be real symmetric and positive definite, where $\sigma_{11}$, $\sigma_{12}$, $\sigma_{22}$ are constants. The coefficients $\sigma_{13},\, \sigma_{23},\, \sigma_{33}$ are functions depending only on $x_1,x_2$ that are determined later. 
The core $D\times\RR$ is insulated, and the exterior region $\RR^3\setminus(\overline{ \Omega}\times\RR)$ is occupied by a homogeneous material with conductivity $\bm{\sigma}_0$ which is possibly anisotropic and of the form
$$\bm{\sigma}_0=
\left( \begin{array}{ccc}
\sigma_{0,11} & \sigma_{0,12}  &0\\
\sigma_{0,12} & \sigma_{0,22} & 0\\
0 & 0 & \sigma_{0,33}\end{array} \right).
$$
The electric potential associated with the described conductivity profile is governed by the equation 
\beq\label{eqn:trans3D}
\begin{cases}
\ds\nabla\cdot\bm{\sigma}_0\nabla\varphi_0=0\quad& \mbox{in }\RR^3\setminus(\overline{\Omega}\times\RR),\\
\ds\nabla\cdot\bm{\sigma}\nabla\varphi = 0 \quad&\mbox{in }\Theta\times\RR,\\
\ds\left(\bm{\sigma}_0\nabla\varphi_0\right)\cdot (n_1,n_2,0)=\left(\bm{\sigma}\nabla \varphi\right)\cdot (n_1,n_2,0)\quad&\mbox{on }\p \Om\times\RR,\\
\ds\varphi_0=\varphi\quad& \mbox{on }\p \Om\times\RR,\\
\ds \left(\bm{\sigma}\nabla \varphi\right)\cdot (n_1,n_2,0)=0\quad&\mbox{on }\p D\times\RR,
\end{cases}
\eeq  
where $\varphi_0$ and $\varphi$ are potential functions in $\RR^3\setminus(\overline{\Omega}\times\RR)$ and  in $\Theta\times \RR$, respectively, and $\Bn=(n_1,n_2)$ is the unit outward normal vector either to $\p \Omega$ or to $\p D$. 
We set $\Be=(e_1,e_2,e_3)$ and $\mathbf{j}=(j_1,j_2,j_3)$ to be the electric field and its associated current field in the coating phase $\Theta\times\RR$. The constitutive relation between them is
\begin{equation}\label{eqn:constitutive}
\nabla\cdot\mathbf{j}=0,\quad \mathbf{j}=\bm{\sigma}\Be,\quad \Be=-\nabla\varphi.
\end{equation}

Our aim is to construct a pair of simply connected domains $(\Omega, D)$ and the coating phase conductivity $\bm{\sigma}$ such that $(\Omega, D)$ is neutral to an applied linear potential $\varphi_0$, i.e., equation \eqnref{eqn:trans3D} admits a solution of which $\varphi_0$ is a linear function.
We write, for ease of notation,
\begin{align}
\varphi_0(x_1,x_2,x_3)&=\psi_0(x_1,x_2)+d_3x_3\label{eqn:assum:phi0}\\
&=d_1 x_1 +d_2 x_2+d_3 x_3 \label{eqn:d1d2}.
\end{align}
In view of the cylindrical structure of the coating phase, we assume
\beq\label{eqn:assum:phi}
\varphi(x_1,x_2,x_3)= \psi(x_1,x_2)+d_3 x_3 \quad\mbox{for some function }\psi.
\eeq

Let us now apply a linear transformation to simplify the problem.

\subsection{Simplification via a linear transformation}
Let $\underline{\mathbf{M}}$ denote the first $2\times2$ submatrix of a $3\times3$ matrix $\mathbf{M}$.
From the assumption on $\bm\sigma$, the submatrix $\underline{\bm\sigma}$ is a constant real symmetric positive-definite matrix.  Hence, it admits a singular value decomposition
$$\underline{\bm\sigma}=\bU \bD \bU^T,\ \bD= \left( \begin{array}{cc}
\lambda_1&0 \\
0& \lambda_2\\
 \end{array} \right),\ \bU^T\bU=\bU\bU^T=\bI_2 $$
for some constants $\lambda_1,\lambda_2>0$ and a constant orthogonal matrix $\bU$. Here, $\bI_2$ denotes the $2\times2$ identity matrix. 
We define a linear transformation $F:\RR^3\rightarrow\RR^3$ as $F(\Bx)= \bF \Bx$,
where the Jacobian matrix $\bF$ is
\beq\label{def:F2_2}
\bF = \left( \begin{array}{cc}
\underline{\bF}&\bold{0} \\
\bold{0}& 1\\
 \end{array} \right) 
 \qquad\mbox{with}\quad 
 \underline{\bF}=\left( \begin{array}{cc}
\frac{1}{\sqrt{\lambda_1}}&0 \\
0& \frac{1}{\sqrt{\lambda_2}}\\
 \end{array} \right)\bU^T.
 \eeq
We assume $\det(\bU)=1$ and, thus, $\det(\bF)={1}/{\sqrt{\lambda_1\lambda_2}}$.

\smallskip

We set $\widetilde{\Theta}= \left\{ \underline{\bF}\, \By:\By\in \Theta\right\}$ so that 
$F(\Theta\times\RR)=\widetilde{\Theta}\times\RR$, and similarly define $\widetilde{\Om}$ and $\widetilde{D}$, and write \beq\label{F_transform:varphi}
\widetilde{\varphi}_0:=\varphi_0\circ F^{-1}\quad\mbox{and}\quad
\widetilde{\varphi}:=\varphi\circ F^{-1}.
\eeq
Here, the symbol $\circ$ denotes the composition of functions and $F^{-1}$ is the inverse function of $F$.
Then, by changing variables $\bm\eta=F(\Bx)$, equation \eqnref{eqn:trans3D} becomes 
\beq\label{eqn:trans3D2}
\begin{cases}
\ds\nabla\cdot \widetilde{\bm\sigma}_0\nabla\tphi_0=0\quad& \mbox{in }\RR^3\setminus( \overline{\widetilde{\Omega}}\times\RR) ,\\[.5mm]
\ds\nabla\cdot  \widetilde{\bm\sigma}\nabla\tphi = 0\quad&\mbox{in }\widetilde{\Theta}\times\RR,\\[.5mm]
\ds \left(\widetilde{\bm\sigma}_0\nabla\tphi_0\right)\cdot (\tn_1,\tn_2,0)
=\left( \widetilde{\bm\sigma}\nabla\widetilde{\varphi} \right)\cdot (\tn_1,\tn_2,0)\quad&\mbox{on }\p \widetilde{\Om}\times\RR,\\[.5mm]
\ds \tphi_0=\tphi\quad&\mbox{on }\p\widetilde{\Om}\times\RR,\\[.5mm]
\ds  \left(\widetilde{\bm\sigma}\nabla \tphi\right)\cdot (\tn_1,\tn_2,0)=0\quad&\mbox{on }\p \widetilde{D}\times\RR
\end{cases}
\eeq 
with 
\begin{align*}
\ds\widetilde{\bm\sigma}(\bm\eta)=\left.
\frac{\bF\bm{\sigma}\bF^T}{\det(\bF)}\right|_{\Bx=\bF^{-1}\bm\eta}\quad\mbox{and}\quad
\ds\widetilde{\bm\sigma}_0(\bm\eta)=\left.
\frac{\bF\bm{\sigma}_0 \bF^T}{\det(\bF)}\right|_{\Bx=\bF^{-1}\bm\eta}.
\end{align*}
Here, $\widetilde{\Bn}=(\tn_1,\tn_2)$ denotes the unit outward normal vector either to $\p\widetilde{\Om}$ or to $\p\widetilde{D}$.
We then set $\widetilde{\Be}=(\widetilde{e}_1,\widetilde{e}_2,\widetilde{e}_3)$ and $\widetilde{\mathbf{j}}=(\widetilde{j}_1,\widetilde{j}_2,\widetilde{j}_3)$ to be the electric field and its associated current field in the coating phase $\widetilde{\Theta}\times\RR$. The constitutive relation between them is 
\begin{equation}\label{eqn:constitutive2}
\nabla\cdot\widetilde{\mathbf{j}}=0,\quad \widetilde{\mathbf{j}}
=\widetilde{\bm\sigma}\widetilde{\Be},\quad \widetilde{\Be}=-\nabla\widetilde{\varphi}.
\end{equation}

One can easily derive that
\begin{align}\label{tildesigma}
 \widetilde{\bm\sigma}
 &=\sqrt{\lambda_1\lambda_2}
\left( \begin{array}{ccc}
 1&0&h_1\\
 0&1&h_2\\
 h_1&h_2&\sigma_{33}
 \end{array}\right)
 \quad\mbox{with}\quad
 \left(\begin{array}{cc}
h_1\\
h_2\end{array}\right)=\underline{\bF}
\left(\begin{array}{cc}
\sigma_{13}\\
\sigma_{23}
\end{array}\right),\\[1mm]
 \widetilde{\bm\sigma}_0
&=\sqrt{\lambda_1\lambda_2}
\left( \begin{array}{cc}
\ds{\bm\sigma}_0^{aniso}&\bold{0} \\[1mm]
\ds\bold{0}& \sigma_{0,33}
 \end{array} \right)
 \quad\mbox{with}\quad
  {\bm\sigma}_0^{aniso}
= \underline{\bF}\, \underline{\bm\sigma_0}\, \underline{\bF}^T.
\end{align}

\subsection{Two-dimensional formulation}\label{section:twodimensional}
By defining $\widetilde{\psi}_0$ and $\widetilde{\psi}$ similarly to \eqnref{F_transform:varphi}, we have
\begin{align}
\widetilde{\varphi}_0(\eta_1,\eta_2,\eta_3)&=\widetilde{\psi}_0(\eta_1,\eta_2)+d_3\eta_3 \label{eqn:psi_0:def}\\
&=(d_1\ d_2)\underline{\bF}^{-1}(\eta_1\ \eta_2)^T+d_3\eta_3\label{eqn:psi_0:linear},\\
\widetilde{\varphi}(\eta_1,\eta_2,\eta_3)&=\widetilde{\psi}(\eta_1,\eta_2)+d_3\eta_3.\label{eqn:phitilde}
\end{align}
Then, it is straightforward to see from \eqnref{eqn:constitutive2} that the electric field and the current field in the coating phase satisfy 
$$
\ds(\widetilde{e}_1,\widetilde{e}_2)=-\nabla \widetilde{\psi},\quad \widetilde{e}_3=-d_3
$$
and
\begin{align}
&(\widetilde{j}_1,\widetilde{j}_2)=\sqrt{\lambda_1\lambda_2}\,\Big[(\widetilde{e}_1,\widetilde{e}_2)-d_3(h_1,h_2)\Big],\\
&\widetilde{j}_3 = \sqrt{\lambda_1\lambda_2}\,\Big[(h_1,h_2)\cdot(\widetilde{e}_1,\widetilde{e}_2)-d_3\sigma_{33}\Big].
\end{align}
{
On the basis of \eqnref{eqn:trans3D2} and \eqnref{eqn:constitutive2}, it can easily be derived that 
\begin{align}\label{eqn:psipair:ver0}
\begin{cases}
\ds\nabla\cdot\left({\bm\sigma}_0^{aniso}\nabla\widetilde{\psi}_0\right)=0\quad&\mbox{in } \RR^2\setminus\overline{\widetilde\Om} ,\\
\ds\Delta\widetilde{\psi} +d_3\nabla\cdot(h_1,h_2)=0 \quad&\mbox{in }\widetilde{\Theta},\\
\ds \widetilde{\psi}_0=\widetilde{\psi}\quad& \mbox{on }\p\widetilde{\Om},\\
\ds \left({\bm\sigma}_0^{aniso}\nabla{\widetilde{\psi}}_0\right)\cdot \widetilde{\mathbf{n}}=\nabla\widetilde{\psi}\cdot \widetilde{\mathbf{n}}+d_3(h_1,h_2)\cdot\widetilde{\mathbf{n}}\quad&\mbox{on }\p \widetilde{\Om},\\
\ds\nabla\widetilde{\psi}\cdot \widetilde{\mathbf{n}} +d_3(h_1,h_2)\cdot\widetilde{\mathbf{n}} =0\quad&\mbox{on }\p \widetilde{D}.
\end{cases}
\end{align}
}

We further specify the material parameters $(\sigma_{13},\sigma_{23},\sigma_{33})$ with which the problem \eqnref{eqn:trans3D2} admits the two-dimensional formulation in section \ref{sec:Neutral2D}. 
Briefly, our assumptions on the parameters relate to the first $2\times2$ submatrix of ${\bm\sigma}$ and the flux of $\widetilde{\psi}$ on $\p\widetilde{D}$. 
First, we impose the restriction that $(\sigma_{13},\sigma_{23})$ is given by
\beq\label{sigma13sigma23:def}
\left(\begin{array}{cc}
\sigma_{13}\\
\sigma_{23}
\end{array}\right)=\underline{\bF}^{-1}
 \left(\begin{array}{cc}
h_1\\
h_2\end{array}\right),\quad 
(h_1,h_2)=\nabla V,
\eeq
where $V=V(\eta_1,\eta_2)$ is a solution to 
\begin{align}\label{sigma13sigma23:def2}
\begin{cases}
\Delta V=0\quad&\mbox{in }\widetilde{\Theta},\\
\nabla{V}\cdot \widetilde{\mathbf{n}}=0\quad&\mbox{on }\p \widetilde{\Omega},\\
\nabla{V}\cdot \widetilde{\mathbf{n}}=-\frac{1}{d_3}g\quad&\mbox{on }\p \widetilde{D}\end{cases}
\end{align}
for some function $g$ whose integral over $\p \widetilde{D}$ vanishes (zero net flux). This restriction ensures that we can still solve the problem using conformal mappings. We assume $d_3\neq 0$.
{Given $\mathbf{h}=(h_1,h_2)$ satisfying $\mathbf{h}=\nabla V$ with $\Delta V=0$ in $\widetilde{\Theta}$ and $\mathbf{h}\cdot\widetilde{\mathbf{n}}=0$ on $\p\widetilde{\Om}$, one can define $g$ as the value of $-d_3 \mathbf{h}\cdot\widetilde{\mathbf{n}}$ on $\p\widetilde{D}$. Conversely, given a flux $g$ such that there is no net flux through $\p \widetilde{D}$, a unique potential $V$ (neglecting the constant term) exists that satisfies \eqnref{sigma13sigma23:def2}, which determines $\mathbf{h}$.
Thus, determining $g$ is equivalent to determining $(h_1,h_2)$.}  The zero net flux condition on $g$ is necessary for the problem \eqnref{sigma13sigma23:def2} to admit a solution.
 Then, we choose $\sigma_{33}(x_1,x_2)$ so that 
\beq\label{eqn:sigma33}
{\sigma_{33}>h_1^2+h_2^2}
\eeq
which implies positiveness for $\bm\sigma$.
Note that the defined parameters $\sigma_{13},\sigma_{23},\sigma_{33}$ are independent of the variable $\eta_3$.

\smallskip

Assuming \eqnref{sigma13sigma23:def}-\eqnref{eqn:sigma33}, 
$(\widetilde{\psi},\widetilde{\psi}_0)$ satisfies
\begin{align}\label{eqn:psipair}
\begin{cases}
\ds\nabla\cdot\left({\bm\sigma}_0^{aniso}\nabla\widetilde{\psi}_0\right)=0\quad&\mbox{in } \RR^2\setminus\overline{\widetilde\Om} ,\\
\ds\Delta\widetilde{\psi}=0 \quad&\mbox{in }\widetilde{\Theta},\\
\ds \widetilde{\psi}_0=\widetilde{\psi}\quad& \mbox{on }\p\widetilde{\Om},\\
\ds \left({\bm\sigma}_0^{aniso}\nabla{\widetilde{\psi}}_0\right)\cdot \widetilde{\mathbf{n}}=\nabla\widetilde{\psi}\cdot \widetilde{\mathbf{n}}\quad&\mbox{on }\p \widetilde{\Om},\\
\ds\nabla\widetilde{\psi}\cdot \widetilde{\mathbf{n}} ={g}\quad&\mbox{on }\p \widetilde{D},
\end{cases}
\end{align}
where $\widetilde{\psi}_0$ is a linear function given by \eqnref{eqn:psi_0:linear}. 
The problem \eqnref{eqn:psipair} with the uniformity condition \eqnref{eqn:psi_0:linear} is over-determined such that there exists a solution only for certain pairs of regions $(\widetilde{\Omega},\widetilde{D})$ and ${g}$. 
As shown in section \ref{sec:Gauert}, for a given $2\times 2$ constant real symmetric positive-definite matrix ${\bm\sigma}_0^{aniso}$ and a simply connected domain $\widetilde{\Om}$, we can construct $\widetilde{D}$ such that  $(\widetilde{\Omega},\widetilde{D})$ is neutral to a given uniform field of arbitrary direction with the choice of $g$ depending on the direction of the uniform field. 
{After determining $g$, we can determine $(h_1,h_2)$, or equivalently $(\sigma_{13}, \sigma_{23})$, and then choose $\sigma_{33}$ such that it satisfies \eqnref{eqn:sigma33}.}
As a result, we obtain cylindrical inclusions of non-elliptical shapes in three dimensions:
for a given $\Om$, $\underline{\bm{\sigma}}$, $\bm{\sigma_0}$, and $\varphi_0$ (satisfying the appropriate conditions assumed in the derivation), we can construct a cylindrical inclusion $(\Om\times\RR,D\times\RR)$ with the conductivity $\bm{\sigma}$. This inclusion is neutral to the uniform field $\varphi_0$, where the entries $(\sigma_{13},\sigma_{23},\sigma_{33})$ of $\bm{\sigma}$ are functions of $x_1$, $x_2$ determined to satisfy \eqnref{sigma13sigma23:def}-\eqnref{eqn:sigma33}.

{The parameters $(\sigma_{13},\sigma_{23},\sigma_{33})$ defined by \eqnref{sigma13sigma23:def}-\eqnref{eqn:sigma33} with an arbitrary function $g$ are in general functions depending on the $x_1,x_2$ variables.}
In the case when $(\widetilde{\Om}, \widetilde{D})$ admits a solution for the two-dimensional problem \eqnref{eqn:psipair} with $g \equiv 0$, then $V$ as determined by \eqnref{sigma13sigma23:def2} is a constant. Hence, $(\sigma_{13},\sigma_{23})$ given by \eqnref{sigma13sigma23:def} are zero. Therefore, apart from a possible variation in $\sigma_{33}(x_1,x_2)$,
the corresponding neutral cylindrical inclusion has a shell of constant conductivity.
Conversely, if the shell $(\Om\setminus\overline{D})\times\RR$ has constant conductivity, then $V$ given by \eqnref{sigma13sigma23:def} is a linear function of $x_1$ and $x_2$. 
In fact, the function $V$ has to be constant to satisfy the zero flux condition on $\p \widetilde{\Omega}$ in \eqnref{sigma13sigma23:def2} and hence $g=0$.
The solution shapes $(\widetilde{\Om}, \widetilde{D})$ to equation \eqnref{eqn:psipair} with $g\equiv0$ were previously found \cite{Milton:2001:NCI}. 
In other words, the three-dimensional cylindrical neutral inclusions with constant shell conductivities are those obtained by applying affine transformations to those in  \cite{Milton:2001:NCI}.

\section{Conclusions}\label{sec:conclusions}

This paper presents our constructions of $E_\Omega$-inclusions $D$ in two dimensions based on complex analysis and a conformal mapping from a circular annulus to the domain $\Omega \setminus D$.
Our method does not impose a restriction on the shape of $D$, but generates $E_\Om$-inclusions $D$ with an outer boundary of general analytic shape. The region $\Omega$ needs to be tailored
to avoid singularities in the extended field. By using a similar conformal mapping technique, we also obtain non-elliptical coated inclusions in two dimensions that are neutral if an appropriate flux is applied at the boundary of $D$, and we obtain cylindrical neutral inclusions in three dimensions.

\section*{Acknowledgements}
ML is supported by the National Research Foundation of Korea(NRF) grant funded by the Korea government(MSIT) (No. 2016R1A2B4014530 and No. 2019R1F1A1062782).
GWM is grateful for support from the KAIST mathematics research station(KMRS) and from the National Science Foundation through Grants DMS-1211359 and DMS-1814854.
The authors thank to Hoai-Minh Nguyen for drawing their attention to relevant references.

%
%
 \bibliographystyle{plain}
\bibliography{2019_Lim_Graeme_NNI}     

\end{document}